\documentclass[preprint2, longabstract]{aastex}
\def\kms{\rm{km \ s^{-1}}}
\def\etal{\rm{et al. }}
\def\deg{^\circ }

\clearpage

\begin{document}

\title{The Nature of the Peculiar Virgo Cluster Galaxies NGC 4064 and NGC 4424}
\author{Juan R. Cort\'es} 
\affil{Departamento de Astronom\'{\i}a, Universidad de Chile}
\affil{Casilla 36-D, Santiago, Chile}
\email{jcortes@das.uchile.cl}
\affil{Department of Astronomy, Yale University}
\affil{P.0. Box 208101, New Haven, CT 06520-8101}
\author{Jeffrey D. P. Kenney}
\affil{Department of Astronomy, Yale University}
\affil{P.0. Box 208101, New Haven, CT 06520-8101}
\email{kenney@astro.yale.edu}
\author{Eduardo Hardy
\footnotemark[1]\footnotetext[1]{The National Radio Astronomy Observatory is a facility of the National
Science Foundation operated under cooperative agreement by Associated
Universities, Inc.}}
\affil{National Radio Astronomy Observatory}
\affil{Casilla El Golf 16-10, Las Condes, Santiago, Chile}
\email{ehardy@nrao.edu }
\received{someday} 

\shorttitle{Some title}
\shortauthors{Cort\'es, Kenney \& Hardy}

\begin{abstract}
%%%%%%%%%%% Jeff abstract %%%%%%%%%%%%%%%%%%%%%%%%%%%%%
%%%%%%%%%%%%% Phrase suggested by P. van Dokkum %%%%%%%%%%%%%%
Using extensive kinematical and
morphological data on two Virgo cluster galaxies undergoing
strong nuclear star formation, we show that ram pressure
stripping and gravitational interactions can act
together on galaxies that have recently fallen into clusters.

We present a detailed study of the peculiar HI-deficient
Virgo cluster spiral galaxies NGC 4064 and NGC 4424,
using $^{12}$CO 1-0 interferometry, optical imaging
and integral field spectroscopic observations,
in order to learn what type of environmental interactions
have affected these galaxies.
Optical imaging reveals that
NGC 4424 has a strongly disturbed stellar disk, with
banana-shaped isophotes and shells.
NGC 4064, which lies in the cluster outskirts,
possesses a relatively undisturbed outer stellar disk and a central
bar.
In both galaxies H$\alpha$ emission is confined to the
central kiloparsec, and originates in bar-like strings of
luminous star-forming complexes surrounded by fainter
filaments.
%%%%%%%% Change by Jeff

Complexes of young blue stars
exist beyond the present location of ongoing star formation,
indicating rapidly shrinking star-forming disks.
Disturbed dust lanes extend out to a radius of 2-3 kpc,
much further than the H$\alpha$ and CO emission
detected by us, although similar to the blue stellar complexes.
CO observations reveal bilobal molecular gas morphologies, with
H$\alpha$ emission peaking inside the CO lobes, implying a time
sequence
in the star formation process.
Gas kinematics reveals strong bar-like non-circular
motions in the molecular gas in both galaxies, suggesting that the
material
is radially infalling.
In NGC 4064 the stellar kinematics reveal strong bar-like non-circular
motions
in the central 1 kpc,
and stars supported by rotation with V/$\sigma >$ 1 beyond a radius of
15$"$/1.2
kpc.
On the other hand, NGC 4424 has extremely modest stellar rotation
velocities ($V_{max} \sim$ 30 $\kms$),
and stars are supported by random motions as far out as we can measure
it,
with V/$\sigma$=0.6 at r=18$"$/1.4 kpc.
The ionized gas kinematics in the core are disturbed and possibly
counterrotating.
The observations suggest
that the peculiarities of NGC 4424 are the result of
an intermediate-mass merger plus ram pressure stripping.
In the case of NGC 4064, the evidence suggests
an already stripped ``truncated/normal'' galaxy that recently suffered
a
minor merger or tidal interaction with another galaxy.
Observations of the present star formation rate and gas content,
suggest that
these galaxies will become
small-bulge S0s within the next 3 Gyr.
We propose that galaxies with ``truncated/compact'' H$\alpha$
morphologies such
as these are the result of the independent
effects of ram pressure stripping, which removes gas from the
outer disk, and gravitational interactions such as mergers, which
heat stellar disks, drive gas
to the central kpc and increase the central mass concentrations.
Together these effects transform the morphology of these galaxies.

\end{abstract}


\section{Introduction}

Ever since the times of Hubble \& Humason (1931), observations have shown that
the
 galaxy populations in clusters differ from those in the field. Spitzer \&
Baade
(1951) and Gunn \& Gott (1972) suggested that dynamical processes within
clusters might be responsible for transforming the initial galaxy population.
The
discover of the ``morphology-environment'' relationship (Dressler 1980), the
``Butcher-Oemler effect'' (Butcher \& Oemler 1978), and more recently the
results of the MORPHS collaboration (Dressler \etal 1997, Poggianti \etal
1999)  suggest that the environment in cluster of galaxies is
responsible for the morphological evolution of galaxies, with spirals becoming
lenticular and  redder.
Several mechanisms have been proposed as the cause of this evolution,
such as merging (e.g; Hernquist 1992, Naab \& Burkert 2003, van Dokkum \etal 1999), ICM-ISM stripping
(Gunn \& Gott 1972;
Nulsen 1982; Schulz \& Struck 2001; Vollmer \etal 2001; van Gorkom 2004, Kenney \etal 2004), and
 high speed tidal interactions (also called ``harrassment'' Moore \etal 1996).
However, it is not yet clear which processes  are the main drivers of cluster
 galaxy evolution. Merging has been shown to be
important in the formation of large ellipticals such as cD galaxies, boxy and
disky
 elliptical galaxies, and disturbed spirals (Bournaud \etal 2004).
It has been argued that the high
relative velocities between galaxies in virialized clusters make merging an
unlikely mechanism (Ghigna \etal 1998).
 However,  mergers can occur in the unrelaxed part of clusters where the local
velocity
dispersion may be much lower than that within the virialized inner portion.
Conversely, ICM-ISM stripping, would be effective in removing the ISM from
galaxies in the virialized portion of the cluster.
In fact, under some conditions of ICM density and relative velocities, ram
pressure stripping could be strong enough as to {\it completely} remove the
gas.  
High
 velocity tidal interaction or ``harassment'' (Moore \etal 1996) may make  
faint
Sc-Sd galaxies evolve into dwarf galaxies by tidally removing the
stars, although the disks are not completely destroyed (Mastropietro
\etal 2004).
In the case of Sa-Sb galaxies, their concentrated potential help them
maintain their stability under harassment, but this process could
disturb their disks and increase the scale
height of the stars (Moore \etal 1999). Finally, there is recent
evidence that galaxy transformation is not only restricted to core of the
clusters.
Simulations show that Sc/Sd galaxies under hierarchical growth and
galaxy harrassment undergo a significant transformation even in the outskirts
of the clusters (Mastropietro \etal 2004).
Recent studies find galaxies with
reduced star formation far from the centers of some clusters (e.g.
Balogh et al 1998, Lewis et al 2002, G\'omez et al 2003, Nichol 2004).
Solanes \etal (2001) find HI deficient galaxies located
out to 2 Abell radii in the Virgo cluster.
These may be galaxies that were locally stripped at surprisingly large
cluster distances, or stripped in the core then
rebounded to large clustercentric distances.

The Virgo cluster is the nearest moderately rich cluster (distance of $\sim$ 16
 Mpc, Jacoby \etal 1992; Freedman \etal 1994, yielding a scale of 4.6
kpc/arcmin), with about 2000 identified galaxies Bingelli \etal (1987), of
which
 about 110 spirals and lenticulars are brighter than $0.1 L_{*}$.
Virgo has many disturbed and HI-deficient galaxies, making it  
an ideal place to study in detail the processes that affect galaxy evolution.
The cluster has
 a significant population of galaxies characterized by truncated star formation
morphologies, with no H$\alpha$ in the outer disk but strong H$\alpha$  in the
inner region (Koopmann \& Kenney, 2004). Most of these ``truncated/normal''
galaxies are consistent with having simply ICM-ISM stripped disks ,
 but some of them are more peculiar than that.
  NGC 4064 and NGC 4424 are the prototypical  
members of  a type of H$\alpha$ radial distribution which Koopmann \& Kenney
(2004) describe as ``truncated/compact''. Compared to  typical isolated
galaxies, ``truncated/compact'' galaxies  have circumnuclear normalized star
formation rates enhanced by a factor of
at least 5, and virtually no star formation beyond the central kiloparsec
(Koopmann \& Kenney 2004). In
contrast to truncated/normal galaxies, these truncated/compact galaxies have
unusually steep H$\alpha$ surface brightness profiles resembling
the stellar profile of a ``bulge''.

 These two peculiar 0.25 L$_{*}$ Virgo cluster spiral
galaxies are particularly interesting because they have small bulges, large HI
deficiencies, and strong H$\alpha$
 emission confined to their cores, with bar-like morphologies. Furthermore, one
of them, NGC4064, is located in the outskirts of the cluster, which may suggest
that its HI deficiency is the result of environmental effects operating in the
outer parts of Virgo. It has been reported that these two galaxies have  H$
\alpha$ rotational gas velocities which are extremely low with respect to other
Virgo cluster galaxies with roughly the same luminosity (Rubin \etal 1999).
While the galaxies have some similarities, their stellar morphologies
differ. NGC 4064 has an outer stellar disk which
contains very open stellar spiral arms but appears relatively undisturbed,
whereas NGC 4424 has heart-shaped optical isophotes,
shells and complex stellar structures which suggest a merger event (Kenney
\etal
1996).

In this paper, we present a detailed studied of the kinematics and morphologies
of
 these two Virgo cluster galaxies  in the hope of identifying the
 physical mechanisms responsible for their peculiarities.  We present a
 comparison between the stellar, ionized gas and molecular gas morphology and
kinematics in the form of two-dimensional maps. In \S 2, we briefly describe
the
known properties of these galaxies.
 In \S 3 we present the technical details of our spectroscopic observations as
well as those of the CO interferometry and optical imaging.
The observational  results related to
 optical morphology, molecular gas morphology, and kinematics, dust morphology,
stellar, and ionized gas kinematics are given in \S 4.
We then discuss the
 nature of these objects in \S 5, and summarize our results in \S 6.

\section{The Galaxies}

\subsection{NGC 4064}

\begin{figure}
\plotone{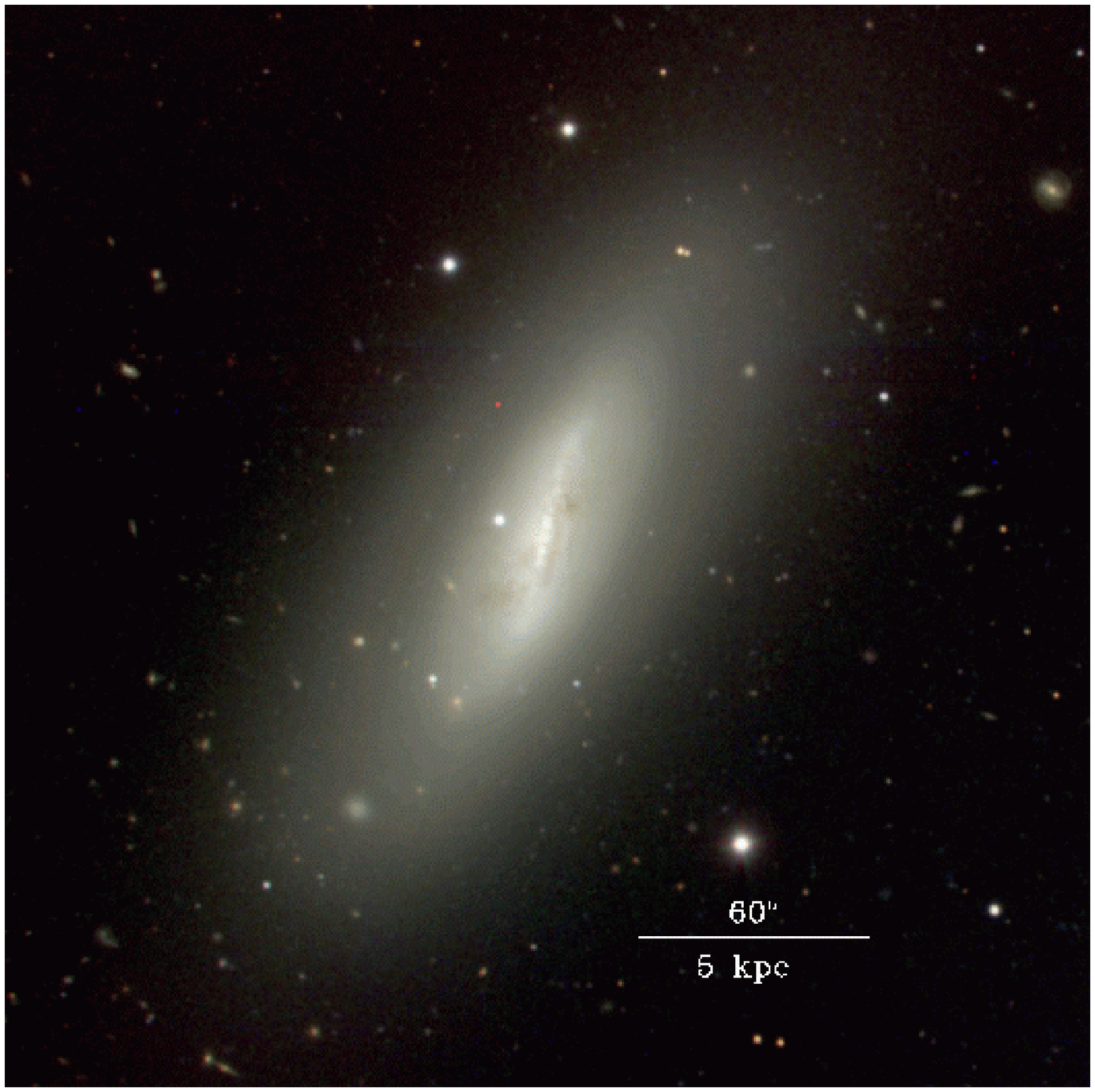}
\caption{Composite BVR WIYN image of NGC 4064. The galaxy exhibits a central
bar, with unperturbed outer stellar disk, and irregular dust lanes in the inner
 2 kpc.}
\label{n4064rgb}
\end{figure}

NGC 4064 is a peculiar SB galaxy located 8.8$\deg$ (2.5 Mpc) from
M87, in the outskirts of the Virgo cluster, and is among the
most distant known cluster members. The optical image
shows perturbed dust lanes and a central bar ending in very open
spiral arms (Fig. $\ref{n4064rgb}$).
H$\alpha$ imaging by Koopmann \etal (2001) shows that the H$\alpha$ emission
is confined to the central kpc.  The H$\alpha$
surface brightness in this central region is among
the highest in the Virgo Cluster (Koopmann \& Kenney 2004).
NGC 4064 has been reported to have
low H$\alpha$ line-of-sight velocities for its luminosity with $V_{max} \sim
$40
$\kms$
(Rubin \etal 1999), but has a HI linewidth of 163 $\kms$ (Helou \etal
1984) consistent with its luminosity.
It has a large HI deficiency of H$_{def}$= 1.0, corresponding to one-tenth
the HI for a normal spiral of its luminosity.
A table of general galaxy properties is given in Table 1.

\subsection{NGC 4424}
\begin{figure}
\plotone{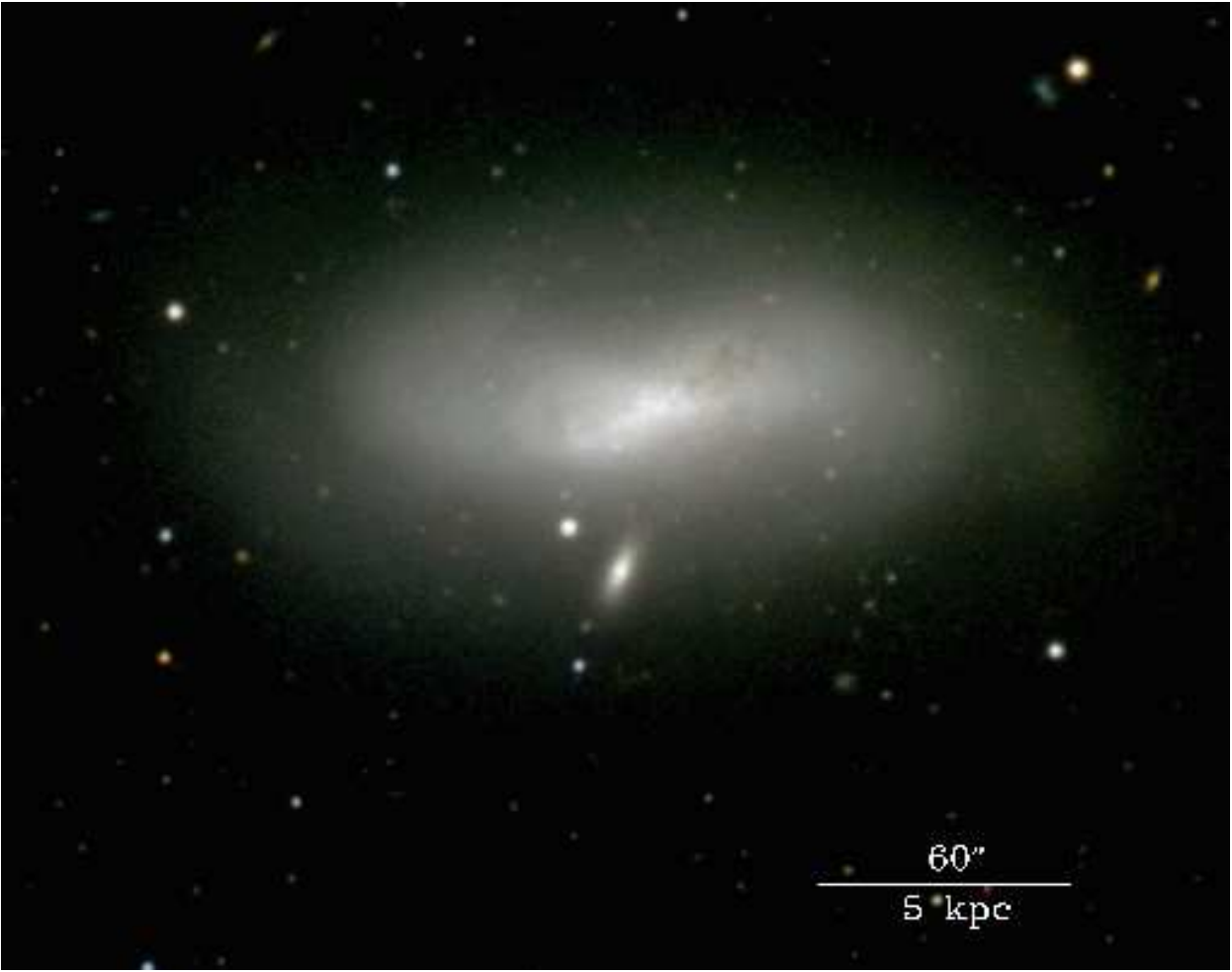}
\caption{Composite BVR WIYN image of NGC 4424. The galaxy present a disturbed
 stellar disk with shell-like features, and banana-shaped isophotes. Irregular 
dust lanes are present in the inner 3 kpc.}
\label{n4424rgb}
\end{figure}

NGC 4424 is a peculiar Sa galaxy located  3.1$\deg$ (870 kpc projected
distance) from M87. It shows a more disturbed morphology than NGC 4064,
characterized by unusual heart-shaped inner R isophotes
and shell like features (Fig. $\ref{n4424rgb}$), suggesting a merger
event (Kenney \etal 1996).
It also has H$\alpha$ emission confined to the central  kpc and one of the
highest H$\alpha$ surface brightnesses within the Virgo cluster (Koopmann \&
Kenney 2004). H$\alpha$ velocities derived by Rubin \etal (1999) are extremely
low ($\sim$ 40  $\kms$).
Its HI deficiency is also high ($\sim$ 1.0) (Helou etal 1984; Kenney
\& Young 1989).

\section{Observations and Data Reductions}

\subsection{Optical Imaging}
We refer the reader to Table 2 for the  details of the instrumental setup and
observing conditions.
NGC 4064 and NGC 4424 were observed using the Mini-Mosaic imager (Saha \etal
2000) on the WIYN 3.5-meter telescope on Kitt Peak
during 2002 March 12-13 and 2001 March 15, respectively.
The Mini-Mosaic imager has a 4096$\times$4096 pixels detector,
with a 10' field-of-view and a pixel scale of 0.14". NGC 4064  was observed in
B, V, R, and the W15 H$\alpha$ narrow band filter.  
For each filter three exposures were taken, with integration times of
180 s each for the broadband filters and 300s each for the narrowband filter.
Images were taken in a dither pattern, with incremental offsets of
12" North-South and 5" East-West, in order to help fill in the
gap between the two CCDs of Mini-Mosaic.
In the case of NGC 4424, only the filters R and W15
 (H$\alpha$) were used, with three exposures of 300 s and three exposures
of 500 s taken through these filters respectively.
B, and R-band images were obtained using the  S2KB imager at  
WIYN on the night of 1997 May 2. The S2KB imager
employs  a detector of 2048$\times$2048 pixels, with a field of view of 6.8'
and
a pixel scale of 0.2". Three 180 s exposures were taken through each filter.

Both galaxies (Fig. $\ref{n4064images}$) were reduced in a similar
way, using IRAF and
especially developed
IDL\footnotemark[1]\footnotetext[1]{http://www.rsinc.com}
routines kindly provided by A. Saha. First, all raw
 images were corrected by the ``ghosting from saturation in another amplifier''
effect
 using the IDL routine preproc3. Overscan and trimming was done with the
routine ccdproc in the  IRAF package ``mscred''. The zeros and flatfield images
were processed as usual. Following the flatfield correction, the images were
corrected by the difference of the level of each amplifier. The sections
corresponding to the
 same CCD were merged, and the two CCDs were joined into a single image.
 Next, the independent images were co-added in order to obtain a resulting
deeper image.  Finally, H$\alpha$ continuum subtracted and B-R images were
built
from the co-added images.

\subsection{Millimeter-wave Interferometric Observations}

NGC 4064 and NGC 4424 were observed in the $^{12}$CO (1-0) line between January
 and March 2002 using the Owens Valley Radio Observatory (OVRO) millimeter
interferometer, with six 10.4 m telescopes in L and E configurations.
The resulting baselines ranged from 30 to 119 m.
The quasar 3C 273 was used as pass band and phase
 calibrator. The average single side-band system temperature was 500 - 600 K. 
The data were calibrated using OVRO's software package MMA (Scoville \etal 
1993). The CO cube
 was mapped and deconvolved using the task IMAGR within the software AIPS. The
resulting spatial resolution was $\sim 4.0"$ (i.e., 310 pc at 16 Mpc) with 
natural weighting. 

\begin{figure}
\plotone{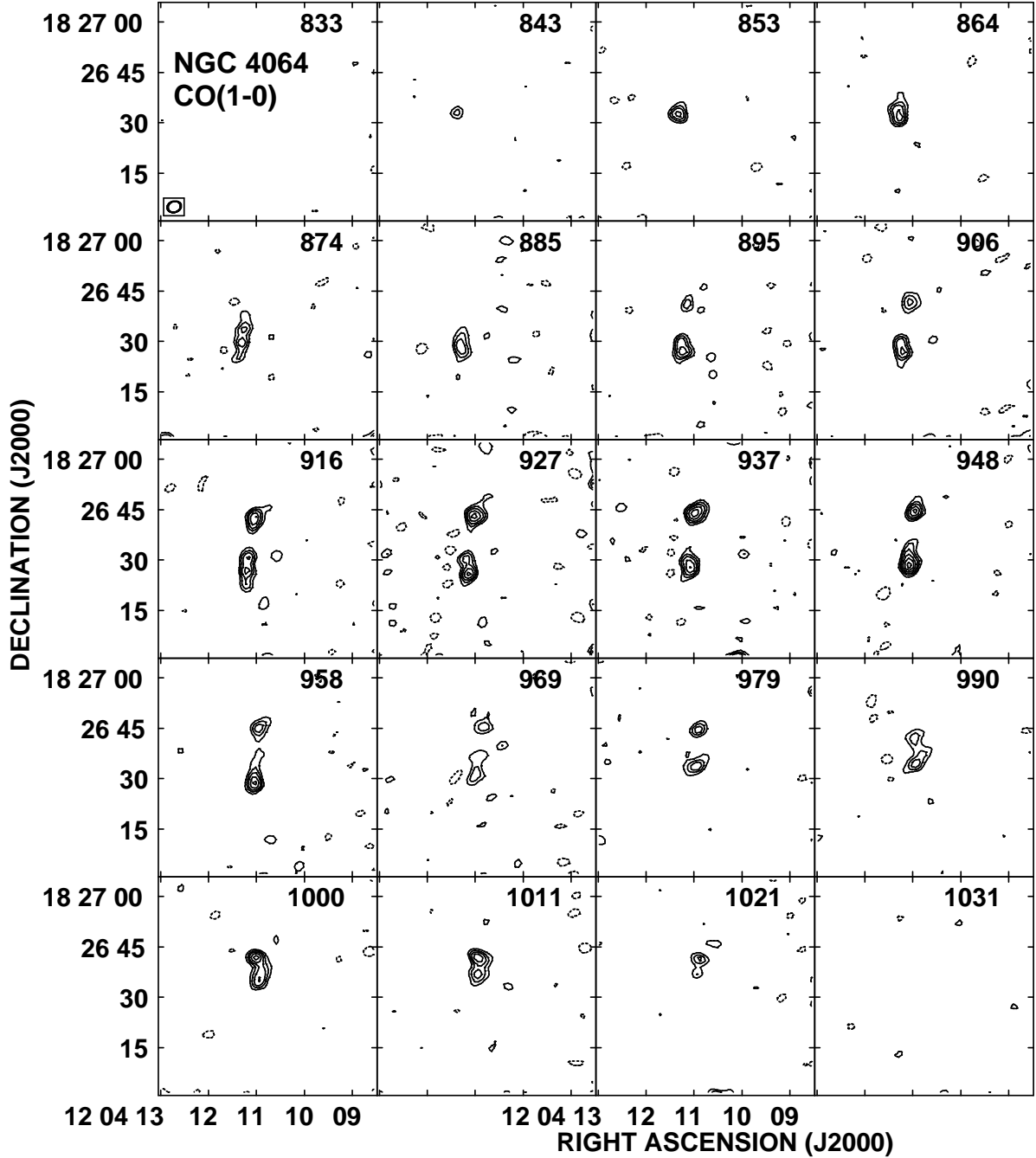}
\caption{OVRO channel maps pf the CO 1-0 line emission for NGC 4064. The spatial 
resolution is 4". The contours are at -3, 3, 4.5, 6, 8, 10, 12, 14, and 16 
$\sigma$, with $\sigma=$ 15 mJy beam$^{-1}$. The heliocentric velocity is given 
in each channel map}
\label{n4064channel}
\end{figure}

\begin{figure}
\plotone{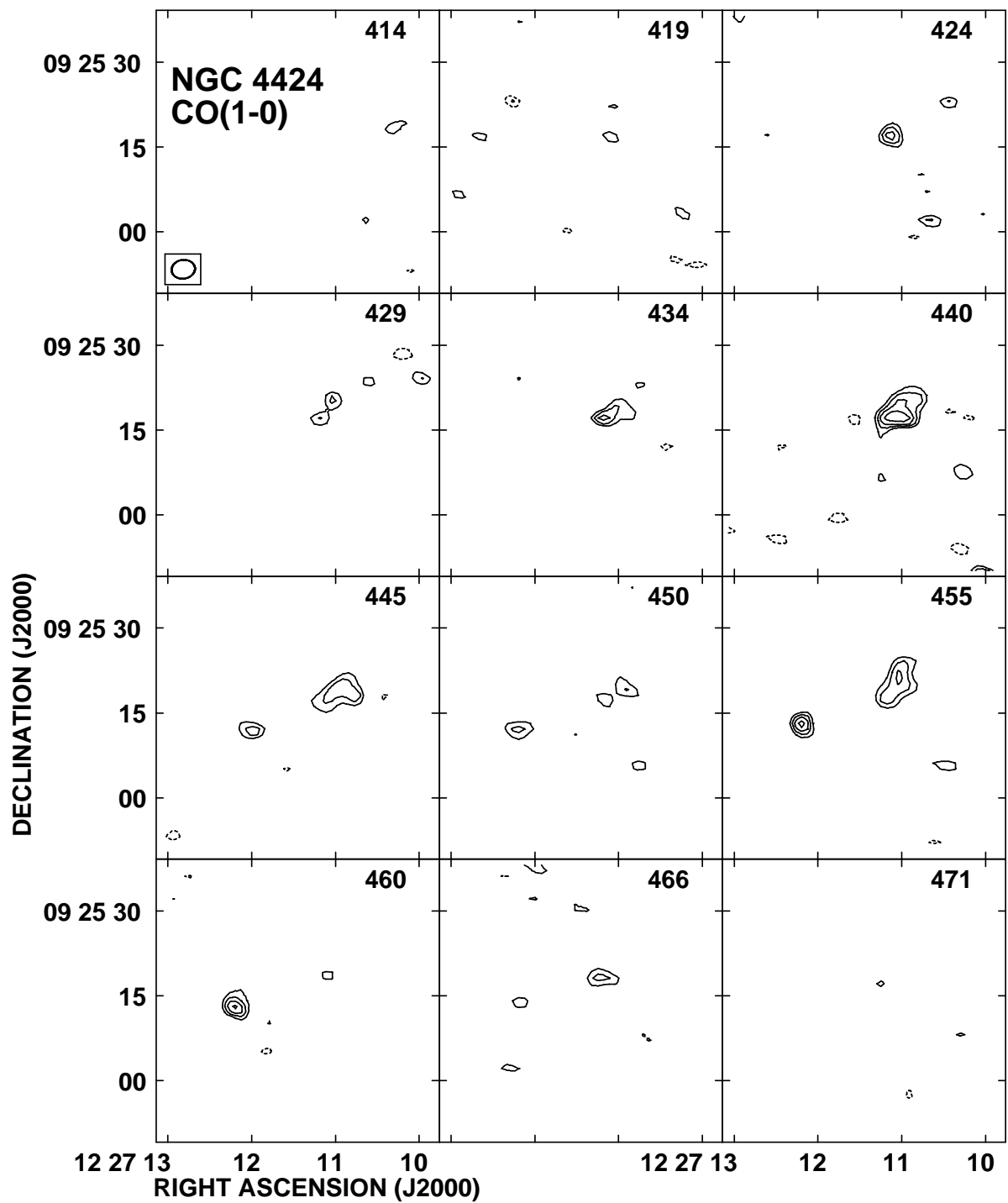}
\caption{OVRO channel maps pf the CO 1-0 line emission for NGC 4424. The spatial
 resolution is 4". The contours are at -4,-3,3,4,5,6,7,8,9,10,11, and 12 
$\sigma$, with
 $\sigma=$ 20 mJy beam$^{-1}$. The heliocentric velocity is given in each 
channel map}
\label{n4424channel}
\end{figure}

For NGC 4064, four spectrometer modules were used resulting in 120 channels with
a spectral resolution of 10.4 $\kms$. The resulting noise per channel was 15 mJy 
beam$^{-1}$, and CO emission was detected ($\geq$ 3$\sigma$) in 18 channels 
over the velocity range 843 to 1021 $\kms$ (Fig. $\ref{n4064channel}$). In the 
case of NGC 4424, two spectrometer modules were used over 120 channels with a 
spectral resolution of 5.2 $\kms$. The resulting noise per channel was 20 mJy 
beam$^{-1}$, and CO emission was detected in 9 channels over the velocity range 
of 424 to 467 $\kms$ (Fig. $\ref{n4424channel}$).

\subsection{Integral Field Spectroscopy}
We refer the reader to Table 4 for the  details of the instrumental setup and 
observing conditions.

NGC 4064 and NGC 4424 were observed with the DensePak Integral Field Unit 
 (Barden, Sawyer \& Honeycutt 1998) at the WIYN 3.5-meter telescope 
during 1999 April 8-9. The 860@30.9 grating at 5000 {\AA} was used at 
second
order, covering the 4500 to 5500 {\AA} wavelength range with a spectral 
dispersion of 0.48 {\AA} per pixel, and a spectral resolution of 2.02 {\AA}. 
Densepak consists of a fiber array of 90 fibers with a  70\%  area coverage, 
each 3.5" in aperture and spaced by 4.0" (300 pc).

The total exposure time in both galaxies was 2 hrs in 4 different
 exposures. The nucleus of the galaxies was centered on the array,
 and the DensePak major-axis was aligned with the optical major axis of the 
galaxy as derived from
 R-Band images by Koopmann \etal 2001. Comparison Cu-Ar lamp exposures
 were taken after every object integration, and G and K giant radial velocities
 standard 
stars were observed with  exposure times of 180 s, for use as template stars for 
the
 kinematical analysis. 

The spectra were reduced with IRAF in the usual way for multifibers 
spectrographs. They were zero--subtracted and overscan--corrected with the 
standard IRAF tasks. Flatfield correction, sky subtraction,
 fiber throughput correction and wavelength  calibration were carried
 out using the IRAF task 
 DOHYDRA in the package HYDRA (Valdes 1995). The galaxy spectra were averaged 
for improving signal-to-noise and cosmic ray removed.

 We derived the two-dimensional stellar velocity field of NGC 4064 and NGC 4424
 using the cross-correlation algorithm ``corrfit'' developed by Statler (1995) 
and
 based on the classic cross-correlation method developed by Tonry \& Davies
 (1979).  The method cross-correlates the galaxy spectra with the template star
 spectra, deriving a cross-correlation function which is the convolution 
between the Line of Sight Velocity Distribution (LOSVD) and the auto-correlation 
function of the template stars.
The LOSVDs is parameterized by a Gaussian although
a more detailed parameterization is also possible via Gauss-Hermite functions 
(van der Marel
 \& Franx 1993), but due to low SNR we didn't attempt such parameterization. 
The spectra were prepared by dividing by the continuum emission, 
rebining logarithmically, and fourier filtering.  
 The emission lines were excluded from the cross-correlation, and the 
template
 star was chosen as the star which gives the highest cross-correlation
 peak with the galaxy spectra.  The outcome of the above process was in the form 
of $V$, $\sigma$ maps. Due to the low signal-to-noise ratio of the
 spectra ($\sim$ 40 per pixel in the central fibers, and $\sim$ 
15 per pixel in the 
outer fibers), the full resolution line-of-sight velocity dispersions
maps were noisy. 
In order to increase the S/N ratio we binned our data using 
the Voronoi 2D binning algorithm of Cappellari\& Copin (2003) and  obtained
 compact bins with a minimum signal-to-noise ratio of 20 per pixel which yielded 
line-of-sight velocity dispersions with a error of 10\% .

In the case of emission lines (H$_{\beta}$) and OIII (5007 A), we derived two-
dimensional ionized
gas velocity maps, by fitting a gaussian function to the emission lines
in the continuum subtracted spectra, obtaining $V$ and $\sigma$ for the ionized 
gas.       

\section{Results}
\subsection{Optical Imaging}
\subsubsection{NGC 4064}
At optical wavelengths, NGC 4064 looks smooth in the outer regions
 beyond 35" (i.e., 2.7 kpc), as shown in Figure $\ref{n4064images}$. 
The outermost isophotes are elliptical with no hint of a  significant 
deviation from ellipticity with $\epsilon \sim$ 0.65, which is consistent with 
an inclination of 70$\deg$. 
The 
ellipticity rises from 0.65 at 35" (2.7 kpc) to 0.72 at 25" (1.9 kpc), over
 the same radial range, the P.A shifts from 155$\deg$ to 165-170$\deg$ due to
 the  presence of a central bar which turns into open spiral arms at about
30" (2.3 kpc). The inner 15" (1.2 kpc) show three main star forming regions, 
separated between them by about 4"  , which are also present in the H$\alpha$ 
image. 
 The R-band profile shows (a) a compact component in the central 10" which is 
partly due to the luminous star forming complexes;
(b) a secondary component between 10" and 35", which coincides with the region 
where
the ellipticity rises, and it is due to the presence of the central bar, and, (c), 
an exponential
disk component with a scale length of 37"$\pm$ 2" (2.9 $\pm$ 0.2 kpc),
which dominates beyond 35" (Fig. $\ref{n4064vdisp}$).
 
The H$\alpha$ emission is confined only to the inner 10",
 being aligned with the bar-like structure (P.A $\sim 170^{{\circ}}$).
This emission consists of four bright star forming complexes
 surrounded by a diffuse component. Some faint filaments can be 
identified in our image. The filaments are
roughly perpendicular to the bar on the eastern side, and form an
arc on the western side.
 These filaments are probably the result of ejection of material from the
luminous star forming HII complexes. 

\subsubsection{NGC 4424}

\begin{figure}
\plotone{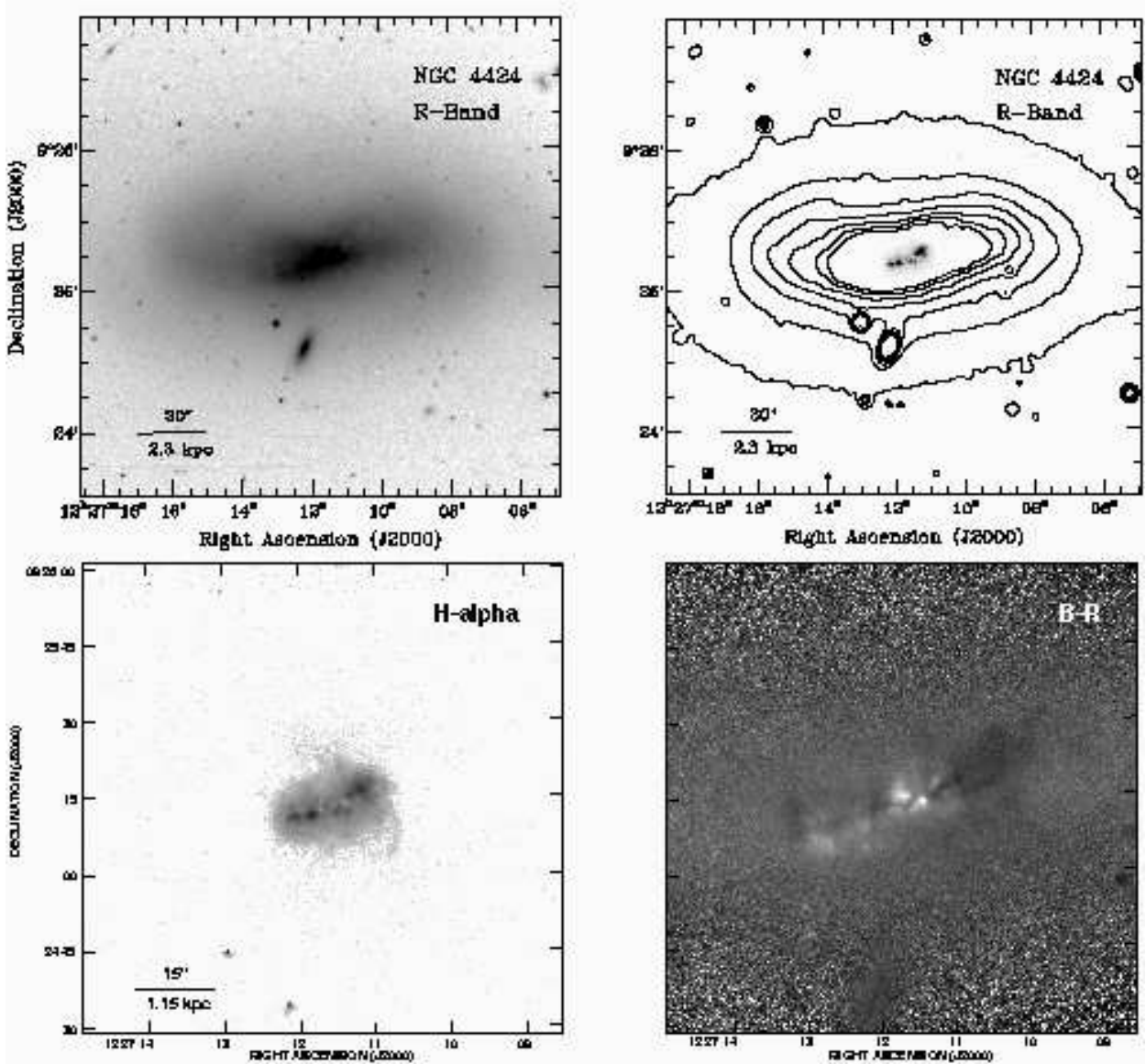}
\caption{R-band and H$\alpha$ images of NGC 4424. Small galaxy 1' south of nucleus is background galaxy. {\em Top Left:} R-band image. 
The inner parts show a disturbed morphology, due to the dust distribution and 
star forming regions, outer parts shows a broad S-shape structure and banana-
shape structures as was noted by Kenney \etal 1996.{\em Top Right:} H$\alpha$ 
continuum subtracted (gray scale), and R-band image (contour lines), the 
H$\alpha$ emission is confined to the inner 1.2 kpc, contours show clearly the 
banana-shape isophotes.{\em Bottom:} H$\alpha$ continuum subtracted image, the 
emission is confined to HII regions along the major axis of the galaxy, 
resembling a bar-like structure or a partial inclinated ring.}
\label{n4424images}
\end{figure}

NGC 4424 shows several stellar non-axisymmetric features
(Fig. $\ref{n4424images}$) previously reported by Kenney \etal 
(1996). The ellipticity is almost constant ($\epsilon \sim$ 0.65)  between 15"
(1.2 kpc) and 70" (5.4 kpc), but the position angle decreases steadily 
from 107$\deg$ to 91$\deg$ due to isophote twisting. In addition 
to this, deviations from ellipticity reveal the presence of unusual
banana-shaped or heart-shaped isophotes at 30" $\leq r \leq$ 70",
characterized by the 
fourier term $s3/a \sim$ 0.05. In the outer parts ($r \geq$ 70"), the
ellipticity is nearly constant ($\epsilon \sim$ 0.47), and the deviations 
from ellipticity are small, but the position angle rises again, reaching 
a maximum of 95$\deg$ at 100" (7.8 kpc) due to the existence of 
  asymmetric broad shell-like features. 
The R-band radial profile
reveals an exponential disk with a small compact bulge-like component 
which dominates in the inner 12"
(Fig. $\ref{n4424vdisp}$). The relatively high ellipticity
($\epsilon \sim$ 0.6), and
diskiness ($ s4 / a \sim$ 0.1) within the inner 10", show that this is a
significantly flattened component, which could be  a circumnuclear disk.

The H$\alpha$ image of NGC 4424 shows strong emission confined to the inner 15"
and originating from several H II complexes which form a slightly curved arc 
centered on the nucleus. The nucleus itself is associated with weak emission,
and the strongest emission is confined to two regions which are 
located symmetrically at 4$"$ (300 pc) from the nucleus, aligned with a P.A of 118$^{{\circ}}$.
Three weak H$\alpha$ filaments are present north of the strong H$\alpha$
emission and oriented perpendicularly to the major-axis. An arc-shaped filament 
is
present in the SW, being probably associated with the western HII complexes. 
 
\subsection{Molecular Gas}

\subsubsection{Morphology}

\subsubsubsection{NGC 4064}

The CO emission in NGC 4064 resembles a bar-like structure
confined to the inner 15" (1.2 kpc), and is mainly
concentrated in two lobes -- as seen in the channel maps of Fig. $\ref{n4064channel}$), spanning the velocity range 843 -- 1021 $\kms$ .
 which are located
nearly symmetrically about the nucleus,
$\sim$ 7" ($\sim$ 540 pc) from the kinematic center
(Fig. $\ref{comoment}$). The total interferometer CO (1-0) line flux is 61 $\pm$ 13 Jy $\kms$,
which is 66 $\pm$ 40\% of the single dish flux
of 93 $\pm$ 40 Jy $\kms$ detected by Kenney \& Young (1988).
Assuming the ``standard'' $N_{H_{2}}/I_{CO}$ conversion factor
(Bloemen \etal 1986) , and a distance of 16 Mpc, the $H_{2}$ mass derived from the
interferometer flux is $M_{H_{2}} \sim 1.7\times 10^{8} M_{\odot}$, and assuming
solar metallicity the total gas mass is 2.3 $\times 10 ^{8} M_{\odot}$.

\begin{figure}
\begin{center}
\epsscale{0.4}
\plotone{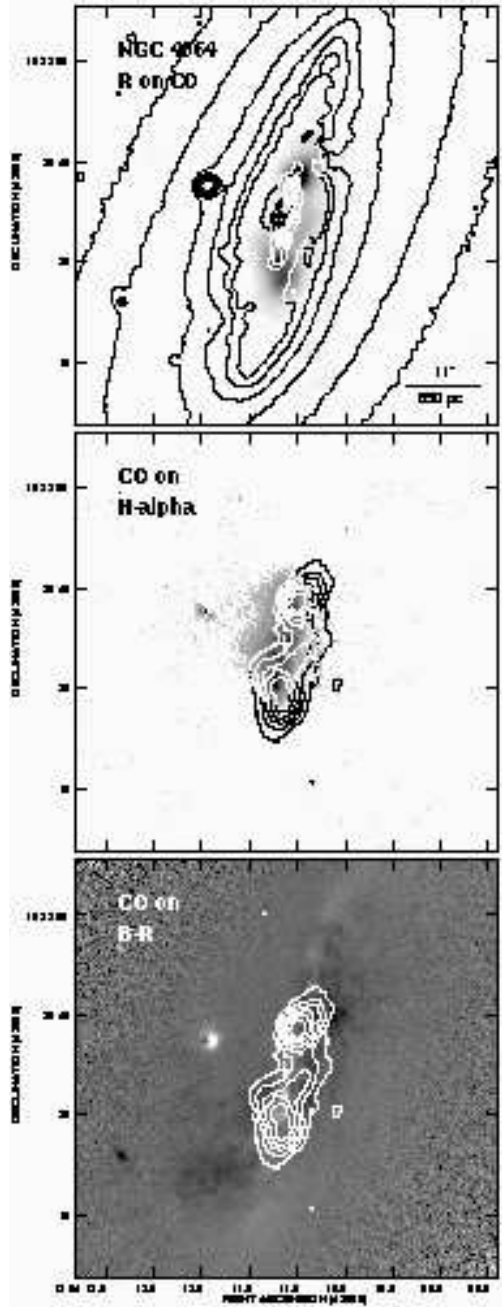}
\end{center}

\caption{NGC 4064 CO emission and optical imaging. {\em Top:} Molecular gas 
(gray scale) and R-band imaging
(contour lines). The bar-like structure is aligned with the CO emission, being 
the CO kinematical center
coincident with the nucleus of the galaxy. 
{\em Middle:} Molecular gas (gray scale) and H$\alpha$ emission
(contour lines). The CO emission is again associated with the H$\alpha$ emission 
complexes, but these
last have their peak of emission slightly displaced to the east.
{\em Bottom:} Disturbed dust lanes (greyscale) and CO emission (contour lines). 
The contour levels are 2, 4, 6, 8, 10, 12, and 14 $\times$ 890 mJy beam$^{-1}$ 
$\kms$. }

\label{picstarco}

\end{figure}

The CO emission is aligned along a P.A of 170$^{{\circ}}$, which is nearly the
same as the stellar bar (Fig $\ref{picstarco}$).
The CO kinematic center coincides with the position of the nuclear stellar
peak.
The CO distribution is asymmetric, about both the major and minor axes.
The southern lobe, with 40 Jy $\kms$, has twice as much
emission as the northern lobe, with 21 Jy $\kms$,
although
the extent of emission along the major axis is similar in the north and south.
%The emission is also very asymmetric about the major axis,
%with 2-3 times more emission to the SW.

CO emission is associated with H$\alpha$ emission, but
slightly displaced ($\sim$ 3"--6"), with the H$\alpha$ peaks located inside in
the CO peaks.
Also, the H$\alpha$ emission is centered on the major axis,
and there is no  H$\alpha$ counterpart to the extra CO extended to the SW.

\subsubsubsection{NGC 4424}

\begin{figure}
\epsscale{0.4}
\begin{center}
\plotone{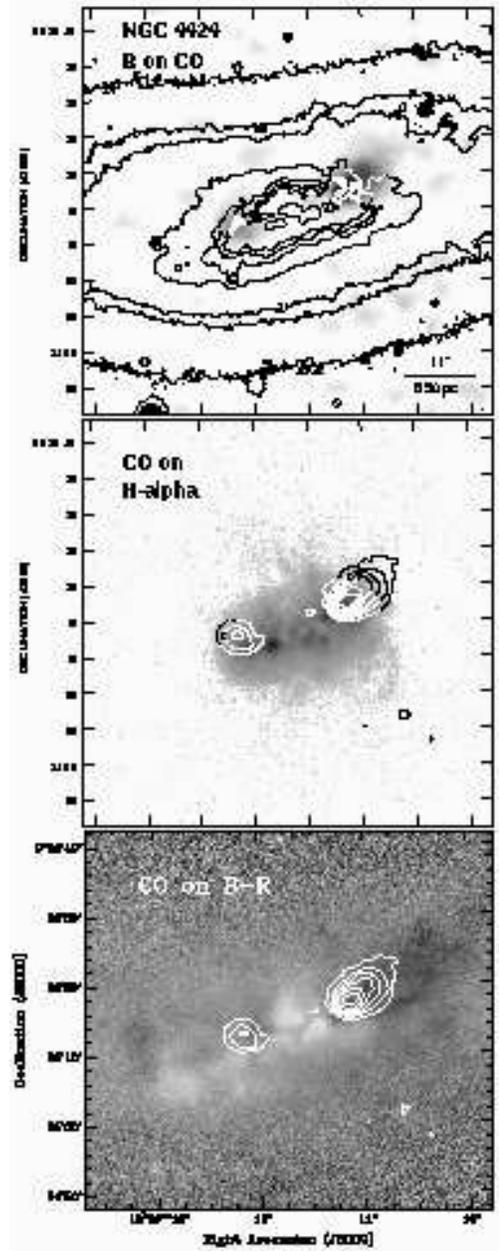}
\end{center}
\caption{NGC 4424 CO emission and optical imaging. {\em Top:} Molecular gas 
(gray 
scale) and R-band imaging (contour lines). The CO structure is not associated 
with any stellar light distribution peak, but the two CO lobes are centered with 
respect to the peak of
the stellar distribution. {\em Middle:} Molecular gas (gray scale)  and 
H$\alpha$ emission (contour lines).
The CO emission is associated with the H$\alpha$ emission, but with the peaks 
slightly displaced.
{\em Bottom:} Disturbed dust lanes (greyscale) and CO emission (contour lines).  
Contour levels are 1, 2, 3, 4, 5, and 6 $\times$ 666 mJy beam$^{-1}$ $\kms$. }
\label{n4424picstarco}
\end{figure}

The CO map of
NGC 4424 shows two lobes of emission (Fig. $\ref{comoment}$), confined
to the inner 15" ($\sim$ 1.2 kpc), as seen in the channel maps of Fig. $\ref{n4424channel}$), spanning the velocity range 424 -- 460 $\kms$.
The lobes are located symmetrically about the
nucleus (Fig. $\ref{n4424picstarco}$),
at radii of 8" ($\sim$ 600 pc), and aligned along
a P.A of 105$^{{\circ}}$, similar to the inner stellar isophotes.
The total CO interferometer flux is 14.3 $\pm$ 2.9 Jy $\kms$,
compared to a single dish flux of 56 $\pm$ 30 Jy $\kms$
(Kenney \& Young 1988), meaning that we are detecting
26 $\pm$ 14 \% of the total CO flux with the interferometer.
Assuming the ``standard'' $N_{H_{2}}/I_{CO}$ conversion factor, and a distance
of 16 Mpc, the $H_{2}$ mass derived from the interferometer flux is
$M_{H_{2}} \sim$ 4.0 $\times 10^{7} M_{\odot}$,
and assuming solar metallicity the total gas mass is
5.5 $\times 10 ^{7} M_{\odot}$.

In addition to the bi-lobal CO morphology,
several other features of the CO morphology in NGC~4424 are similar to NGC~4064.
The CO peaks are associated with the brightest HII regions (Fig.
$\ref{n4424picstarco}$), and
the CO lobes appear offset from the H$\alpha$ peaks by 3"-4",
with the H$\alpha$ peaks inside the CO peaks.
The CO distribution is asymmetric, about both the major and minor axes.
The western lobe, with 10.8Jy $\kms$, has three times as much
emission as the eastern lobe, with 3.5 Jy $\kms$,
although
the extent of emission along the major axis is similar in the east and west.
The emission is also asymmetric about the major axis,
with more emission to the north.
Unlike NGC 4064,
the central HII regions of NGC 4424 are not associated with any detected CO
emission,
although this may be a reflection of the overall weaker CO emission in NGC~4424.

\subsubsection{Kinematics}
\subsubsubsection{NGC 4064}

\begin{figure}
\includegraphics[angle=90,scale=0.8]{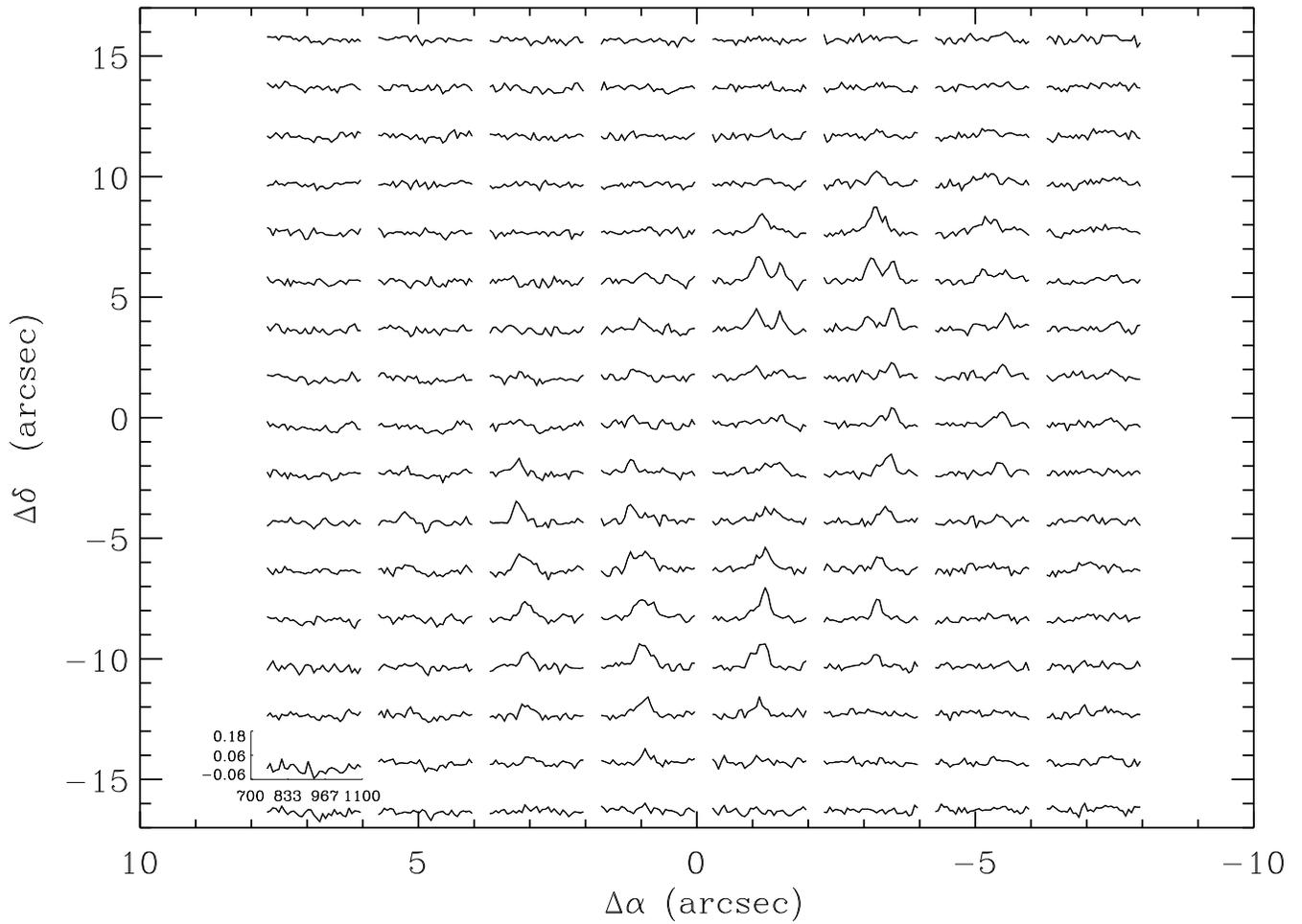}
 \caption{CO(1-0) spectra in NGC 4064 over a region of 10"$\times$17". The 
spectra were average over a 
region of 2"$\times$2". Spacing is 2" and the velocity resolution is 10 $\kms$.
In each spectra, abscissa axis represents the velocity in $\kms$, ordinates axis
represents CO intensity in Jy Beam$^{-1}$}
\label{doublepeak}
\end{figure}

The molecular gas in NGC~4064 has highly non-circular motions,
with bar-like streaming motions and perhaps shocks.
Since the kinematics are somewhat complex, we show channel maps (in (Fig. $\ref{n4064channel}$),
spectral line profiles at 3$''$ resolution (Fig. $\ref{doublepeak}$), position-velocity diagrams,
moment maps, and plots of line-of-sight velocities.

The northern CO lobe shows double-peaked line profiles in some locations,
with distinct peaks near 930 and 1000 $\kms$ (Fig. $\ref{doublepeak}$)
The southern lobe shows broad line profiles,
some of which have complex shapes, although without clear doubling.
A comparison of the line profiles on the 2 sides of the nucleus shows
that the peaks of the broad profiles in the southern lobe
are either associated with one the double peaks in the northern lobe,
or are located between them (after reflection about the systemic velocity).
The total line widths in the 2 lobes are similar, although a bit larger
in the north,
with a maximum FWZI of 165 km/s in the north, and 145 km/s in the south.
The double peaked and broad line profiles are likely due to
shocks associated with a bar.

\begin{figure}
\plotone{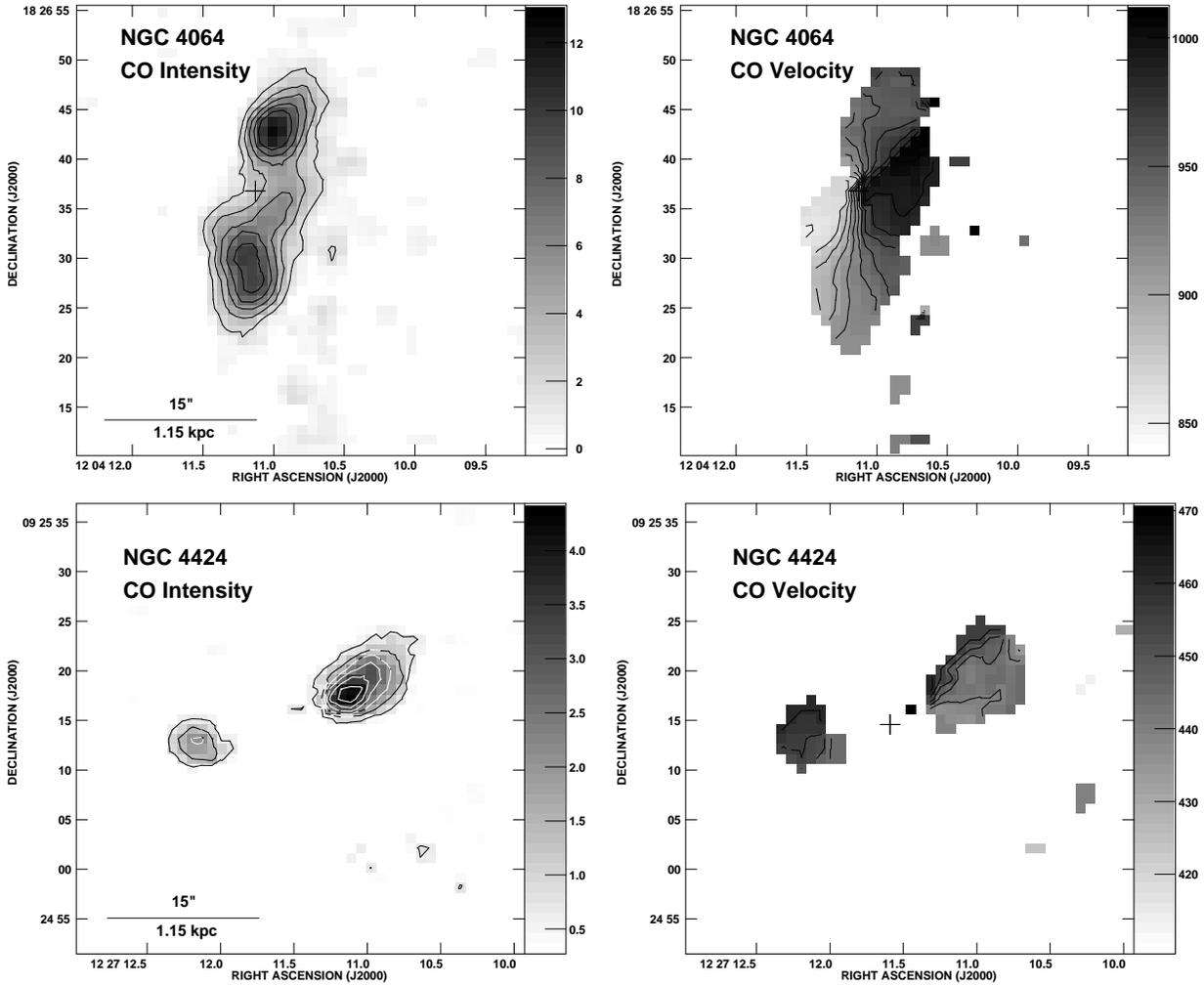}
\caption{Moment maps of NGC 4064 and NGC 4424. The spatial resolution is 4". 
Crosses mark the optical center. {\em Top Left:} NGC 4064 Intensity map. Lowest 
contour is 1780 mJy/beam $\kms$. Contours increments are 2, 4, 6, 8, and 10 
$\times$ 890 mJy/beam $\kms$.  {\em Top Right:} NGC 4064 CO velocity field. 
Contours between 855 and 1005 $\kms$. Contour increment 15 $\kms$. {\em Bottom 
Left:} NGC 4424 Intensity map. Lowest
 contour is 666 mJy/beam $\kms$. Contour interval is 666 mJy/beam $\kms$. {\em 
Bottom Left:} NGC 4424 CO velocity field. Contours between 460 and 440 $\kms$. 
Contour increment 5 $\kms$}
\label{comoment}
\end{figure}

A map of intensity-weighted mean velocities in Figure $\ref{comoment}$
shows a well-ordered velocity field,
with nearly symmetric behavior in the 2 lobes,
and strong non-circular motions (Fig. $\ref{comoment}$).
The CO kinematic major axis varies from $\sim$90$\deg$ in the center,
to $\sim$125$\deg$ at 5$''$, significantly different from
the optical major axis (PA=150$\deg$) and the long axis
of the bar (PA$\sim 170\deg$).
In the lobes the isovelocity contours are nearly parallel to the bar.
While the moment 1 map
shows a representation of the the overall CO velocity field in NGC 4064,
it does not accurately show the kinematics in the region
of double-lined profiles and large velocity gradient,
since we have not spatially resolved the apparent shock.
In positions with double-peaked line profiles, the moment 1 map
shows an intensity-weighted average velocity of the 2 peaks.
Thus the true CO velocity field would show a much larger velocity gradient
in the region of the apparent shock than is shown in the moment 1 map.

\begin{figure}

\plotone{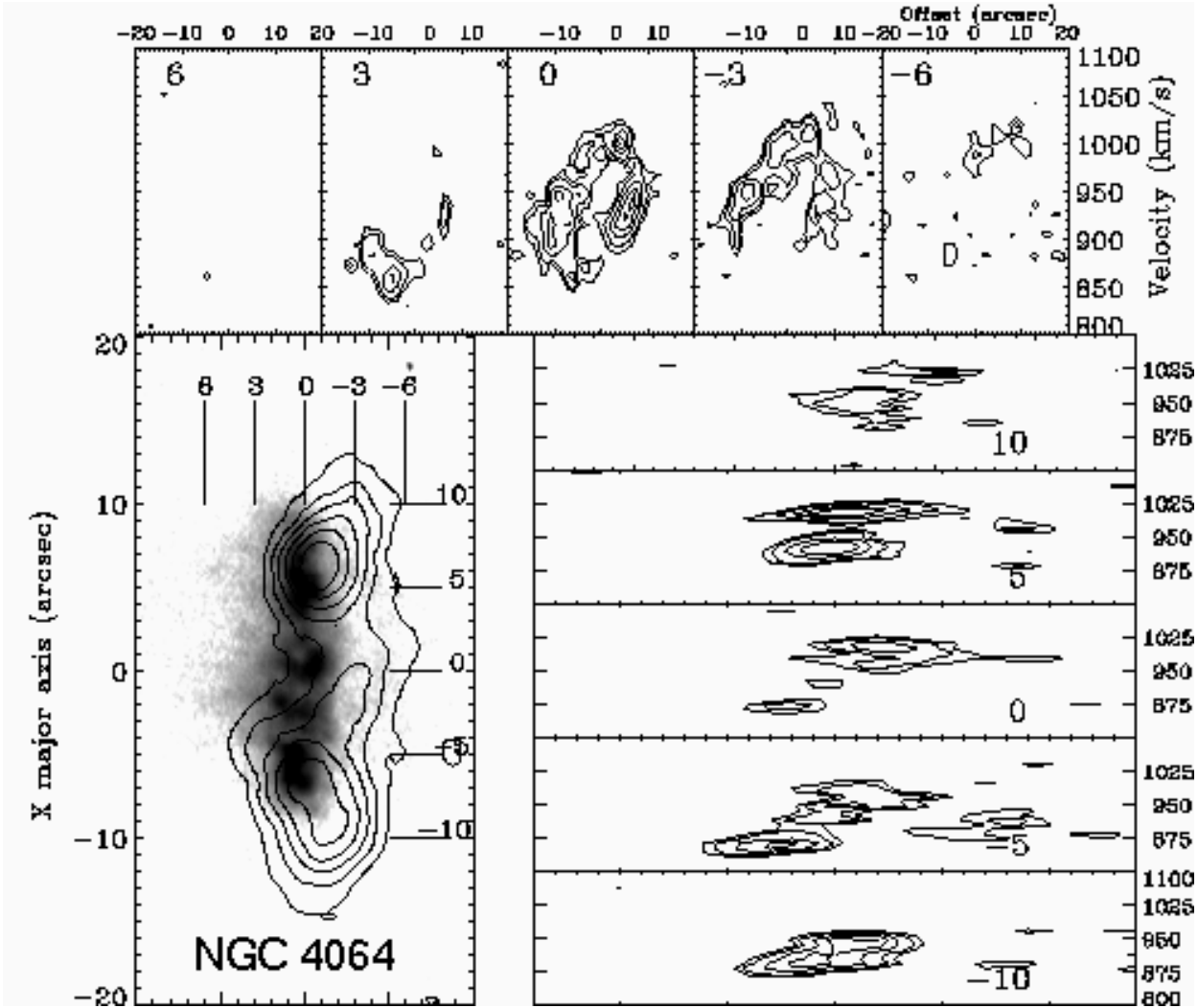}
\bigskip
\bigskip
\caption{Overlay of CO emission intensity on a gray-scale image of the H$\alpha$
 emission of NGC 4064 ({\em left}). CO contours are 2, 4, 6, 8, 10, 12, and 14 
$\times$ 890 mJy
 beam$^{-1}$ $\kms$. CO position-velocity maps were done with a width of 1", 
on different
 regions to different
 offset over the major and minor axis of the bar (P.A = 170$\deg$). Over the 
top,
 we have position-velocity diagrams along the major axis of the bar. They show
 interesting structures, the outermost diagrams show no velocity gradient over 
the
 major axis, but the diagram with offset 0" over the major axis shows emission 
spread out by relatively big range of velocities ($\sim$ 100 $\kms$), revealing 
even 
double peaked emission. The position-velocity diagrams over the minor axis ({\em
right}), reveals 
 significant velocity gradient.}
\label{n4064pvd}
\end{figure}

A truer view of the kinematics in the region of the apparent shock is provided
in Figure $\ref{n4064pvd}$, which shows  position-velocity diagrams parallel and perpendicular
to the stellar bar, which is offset by only 20 degrees from the line of nodes
(major axis).
The PVDs perpendicular to the bar show large velocity gradients
across the CO lobes,
with $\Delta v \sim$150 $\kms$ over 2$''$ (150 pc) in the southern lobe,
and an even larger apparent gradient in the northern lobe.
Similar CO kinematics are observed in other galaxies,
including the barred galaxy NGC 7479 (Laine \etal 1999),
and the Virgo cluster galaxy NGC 4569 (Jogee etal 2004).
In the strongly barred galaxy NGC 7479 (Laine \etal 1999),
CO is strongly concentrated to a narrow ridge along the bar,
with large velocity gradients across the ridge,
presumably due to a shock front.
NGC 4064 also has large velocity gradients perpendicular to the bar.
The broad line profiles in  the southern lobe, and the
double peak profiles in the northern lobe,
could be the result of the
large velocity gradient, marginally resolved with our 4'' resolution,
caused by a shock front.
The higher velocity components (with respect to the galaxy center)
correspond to the preshocked gas,
and the lower velocity components to the postshocked gas.
The difference in shape of the
line profiles between the lobes could be due to a non-uniform
distribution of gas within the velocity field.

The PVDs provide further insight into the asymmetry in line profile shapes
in the southern and northern lobes.
In the PV diagram along the bar through the nucleus (Fig $\ref{n4064pvd}$),
the low level contours show similar curved features on both sides of the  nucleus,
but the high level contours have different distributions.
Whereas in the north most of the CO emission is concentrated at the high and
low velocity ends, with little emission at intermediate velocities.
in the south most of the CO emission peaks at intermediate velocities.
This is consistent with a nearly bi-symmetric velocity field, but a non-uniform
distribution of gas within the velocity field.

Figures $\ref{comoment}$ and $\ref{n4064cut}$,
show that there is a large CO velocity gradient along the minor axis,
totalling $\sim$130 $\kms$ across the central 3$''$.
Given that the southwestern side of the galaxy is the near side (based
on the stellar kinematics (\S4.4.1), and assuming trailing spiral arms),
these correspond to inward radial streaming motions
of 70 $\kms$ in the plane of the galaxy.

\subsubsubsection{NGC 4424}

The CO velocity field in NGC 4424 has similarities with NGC~4064,
although covers less of the central area and is somewhat noiser due to the
weaker CO emission.
The CO isovelocity contours in Figure $\ref{comoment}$ are nearly parallel to the optical
major axis
and the major axis of the CO structure, and nearly perpendicular to
the stellar isovelocity contours (section 4.4).
Both the moment 1 map and the PVDs show a modest CO velocity gradient along
the major axis, with a velocity difference between the 2 CO peaks of only 10 km/s.
There are larger velocity gradients perpendicular to the major
axis, across the short axis of the CO lobes, of
20 km/s across the eastern lobe and 40 km/s across the western lobe.
These motions are consistent with bar-like streaming motions and shocks,
although less extreme than in NGC~4064. Unlike NGC~4064, there is no clear
stellar bar in NGC~4424, and the gas bar may be dynamically decoupled from
the stars.
\subsection{Dust Morphology}
\subsubsection{NGC 4064}
\begin{figure}
\plotone{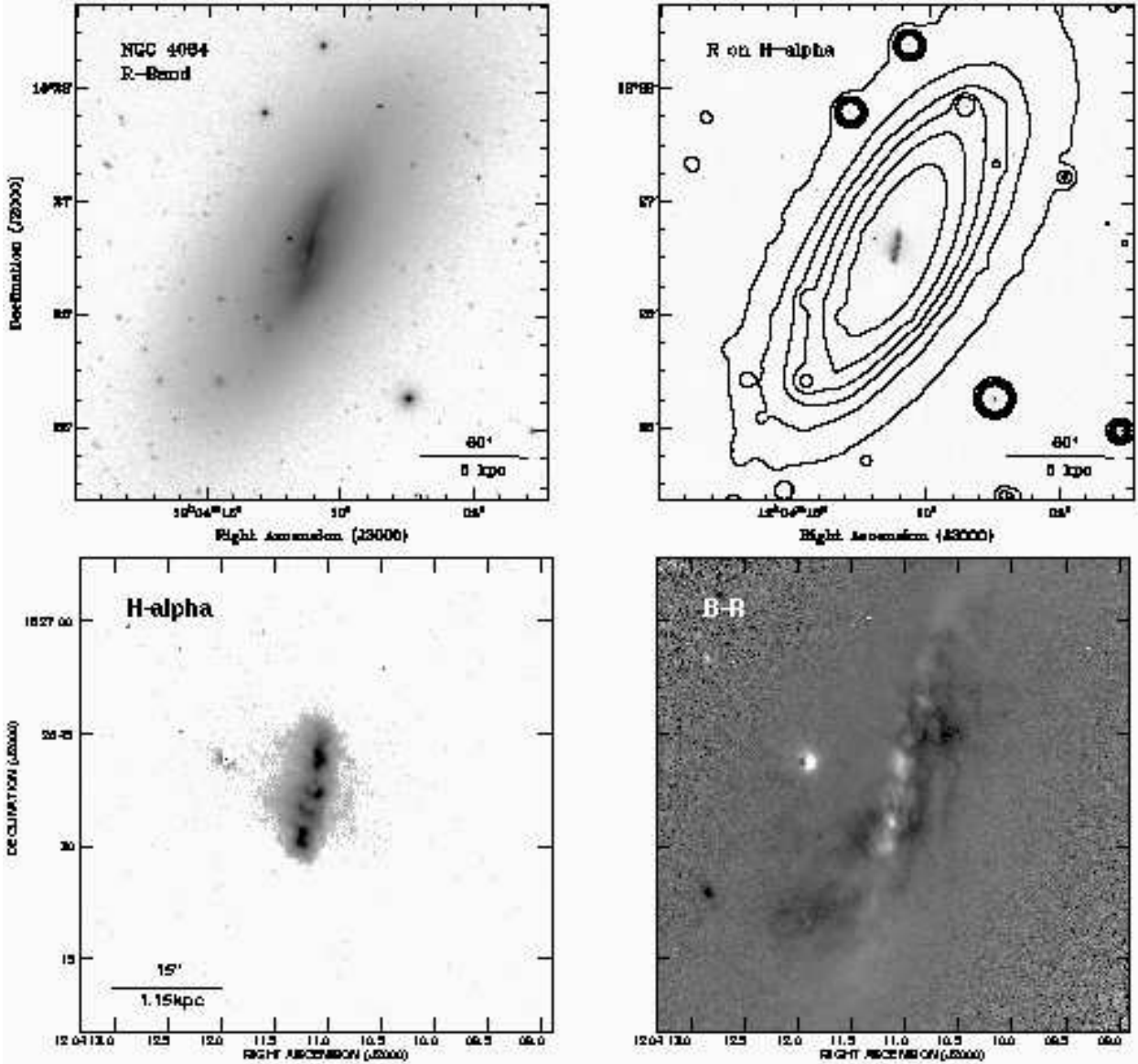}
\caption{R-band and H$\alpha$ images of NGC 4064. {\em Top Left:} R-band image. 
The outer parts of the galaxy look smooth and featureless, an inner bar-like 
structure is revealed with a extension of 30" (2.3 kpc), with a position angle
of 170$\deg$ which is different to the position angle derived from the outer
isophotes (P.A = 150 $\deg$) . {\em Top Right:} H$\alpha$ continuum
subtracted image (gray scale), and R-band image (contour lines).
The H$\alpha$ emission is present in the inner 10" (800 pc) on the bar-like
stellar structure. {\em Bottom Left:} H$\alpha$ continuum subtracted image,
the emission is concentrated in four H$\alpha$ complexes along the bar-like
structure. Filaments are present to the east side of the galaxy, probably
due to ejection of gas due the star formation process.{\em Bottom Right:}
B-R image, disturbed dust lanes are present in the inner 2.3 kpc.}
\label{n4064images}
\end{figure}

NGC 4064 presents a patchy, filamentary and asymmetric dust configuration
 (Fig. $\ref{n4064images}$) very different from the typical 
straight dust lanes found in 
stronger barred galaxies such as NGC 1300, or from the curved dust shapes in 
weak bars as in NGC 1433 (Athanassoula 1992). The B-R map (Fig. 
$\ref{n4064images}$) suggests
more extinction on the western side of the galaxy, implying that it is the near 
side of the disk -- assuming the dust lies on a disk coplanar with the stars.
The B-R image shows the dust is only present in the
 inner 30" (2.3 kpc). There are three main dust concentrations located at
 11" (850 pc) NW of the center, 6" (470 pc) SE, and 18" (1.4 kpc) SE (Fig. 
$\ref{n4064images}$). The two first locations 
 lie over the stellar bar, being connected by an irregular 
 linear feature roughly parallel to the bar. At the third location
the dust is slightly offset from the
 stellar bar. Beyond the inner 10" some of the dust shows filamentary structures 
roughly 
perpendicular to the bar extending to 25" toward the NW and SE.

 The dust lanes extend further from the center of the galaxy
than the H$\alpha$ and CO emission, as expected. The H$\alpha$ emission 
is located in clumps in the inner 15" (1.2 kpc), whereas much of the dust
 surrounds these H$\alpha$ complexes suggesting that the
 disturbed dust distribution near the HII complexes are
 due to the star formation process. 

As expected, dust obscuration appears associated with the CO emission, although it 
is much more widespread and irregular.
Extinction peaks are not perfectly coincident with the CO lobes although the
strongest features are near them. The CO lobes are connected by
 the filamentary dust feature, which also has CO emission associated with it 
(Fig $\ref{picstarco}$). 

The gas distribution associated with the dust is not completely truncated to zero 
at the ``gas truncation radius" (12.5" $\sim$ 1 kpc), defined by the extension of
 the H$\alpha$ emission, probably resulting from environmental effects. The gas 
density is likely to drop 
significantly at the H$\alpha$ gas truncation radius, but there is still enough
 low column density gas beyond it as to have associated patches of dust 
extinction.

\subsubsection{NGC 4424}

NGC 4424 has very disturbed dust lanes (Fig. $\ref{n4424images}$).
The extinction is preferentially concentrated on the north side of the galaxy, 
suggesting that it is the near side of the galaxy.
The B-R image shows dust lanes which extends to $R \sim$ 35" - 40"
(2.7 - 3.1 kpc).  A  filamentary dust structure extends across the
 major axis (P.A = 105$\deg$) of the inner flattened component, or pseudo-bulge. 
This filament survives in the inner 15" (1.2 kpc), ending in two dust
 concentrations ($R = $ 8" or 600 pc) which in turn coincide with the CO lobes 
(Fig. $\ref{n4424picstarco}$). 

The dust concentration associated with the eastern CO lobe is an extension
 of the inner filament. On the other hand, the dust concentration associated
 with the western CO lobe presents a shape similar to the CO lobe,
but is more extended. This concentration continues toward the west in the form 
of two filaments connecting to a patch of dust extending to the north as far as 
$\sim$ 30" (2.3 kpc).
 To the east, a patch of dust is present at $\sim$ 28" (2.2 kpc) from the 
nucleus, and is perpendicular to the major axis of the galaxy.
Weak dust filaments perpendicular to the major axis are visible in the
R-band image within the inner 40" ($\sim$ 3 kpc), although 
they are not evident in the B-R map, probably due to the bigger PSF
of the B image.

As usual the strongest dust absorption coincides with the CO emission.
Since the dust absorption is more extended than the molecular gas, as was the 
case for NGC 4064, probably these
dust lanes are associated with low column density gas that we are not detecting.
 
\subsection{Stellar and Ionized gas kinematics}

\subsubsection{NGC 4064}
\begin{figure}[ht]
\epsscale{1.0}
\plotone{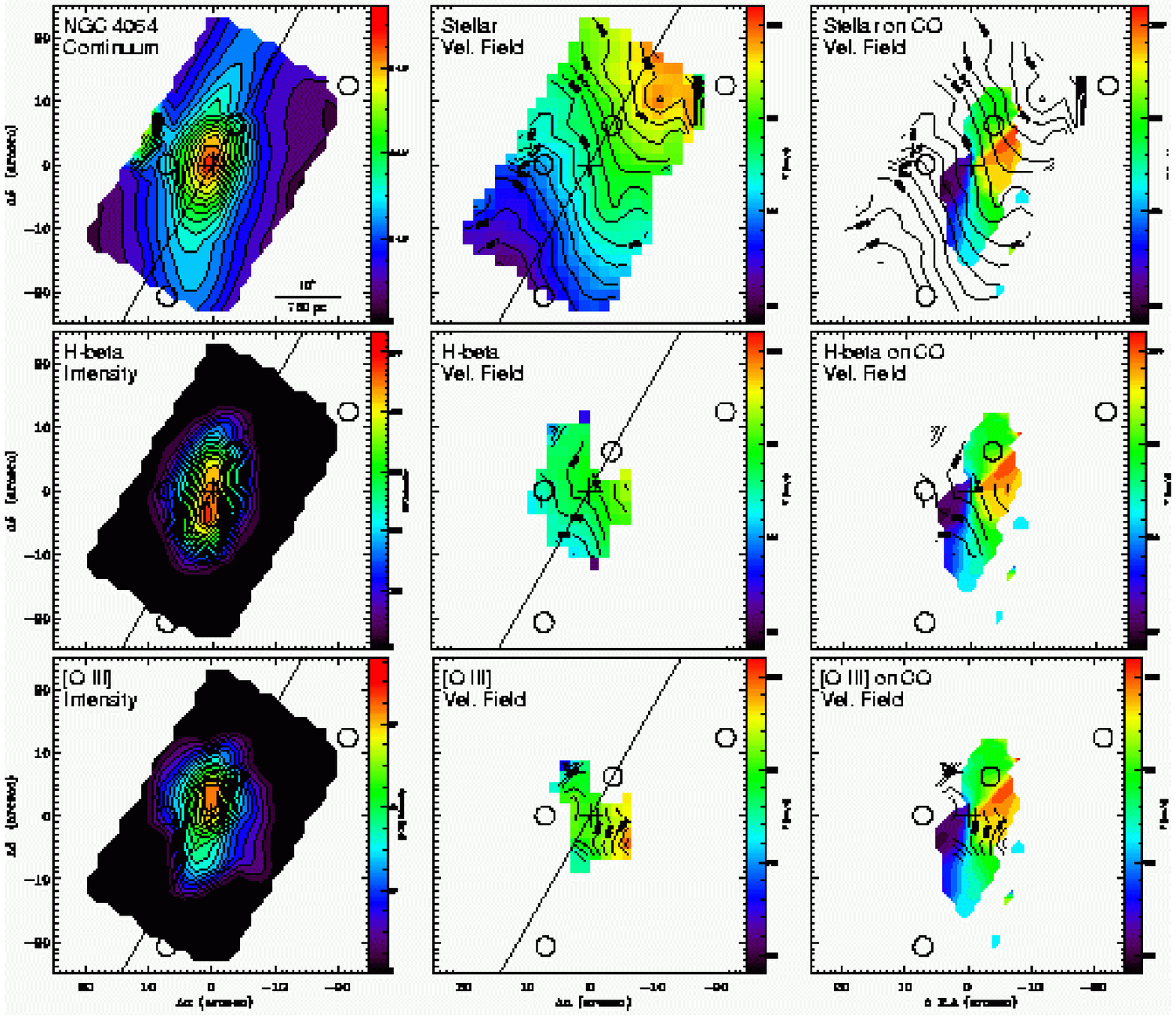}
\caption{DensePak stellar kinematics of NGC 4064. {\em Top Left:} Continuum map. 
{\em Top Middle:} LOS stellar velocity field. {\em Top Right:} LOS stellar 
velocity field (contour lines) over CO velocity field (color) {\em Middle Left:} 
H$\beta$ gas distribution. {\em Middle Middle:} H$\beta$ LOS velocity field. 
{\em Middle Right:} H$\beta$ velocity field (contour lines) over CO velocity 
field (color). {\em Bottom Left:} [O III]$\lambda$5007 gas distribution. {\em 
Bottom Middle:} [O III]$\lambda$5007 velocity field. {\em Bottom Right:} [O 
III]$\lambda$5007 velocity field (contour lines) over CO velocity field (color)}
\label{n4064stellarkin} 
\end{figure}

\subsubsubsection{Stellar kinematics} 
NGC 4064 stellar and ionized gas velocity fields are shown in Fig.
$\ref{n4064stellarkin}$.
 The observed stellar velocity field  is smooth and consistent with rotation 
plus bar-like streaming motions. In the inner 10", the isovelocity contours are 
not parallel to the minor axis, but exhibit the typical S-shape which is a
signature of bar-like streaming motion (Vauterin and Dejonghe 1997).

\begin{figure}
\epsscale{0.6}
\begin{center}
\plotone{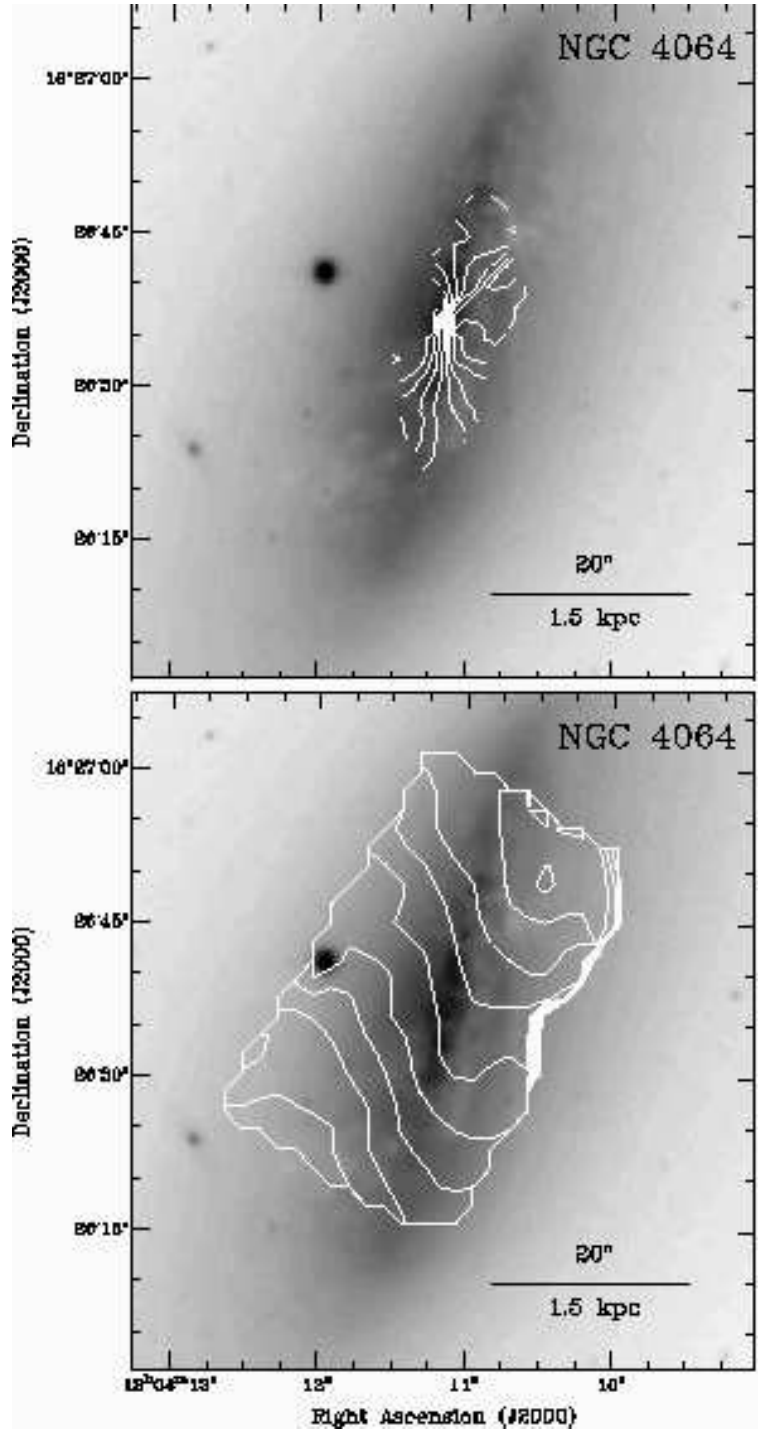}
\end{center}

\caption{CO velocity and stellar velocity, and central bar in NGC 4064. {\em Top:}
R-band image and CO velocity fieldr. Contours run from 860 to 1000 $\kms$ in step
of 10 $\kms$. CO velocity field is roughly perpendicular to the bar. {\em Bottom:}R-band image and stellar velocity field. Contours run from 860 to 1000 $\kms$ in step
of 10 $\kms$. Stellar velocity field exhibit typical S-type shape due to the
influence of the central bar.} 
\label{n4064rbandvel}
\end{figure}

The stellar velocity field in the outer parts of the densepak field of view 
exhibits a velocity gradient almost parallel to the optical major axis of the 
galaxy (Fig. $\ref{n4064stellarkin}$). 
On the other hand, the inner stellar velocity gradient is nearly perpendicular 
to the bar although not as much as the CO velocity gradient (Fig. $\ref{n4064rbandvel}$), a difference 
probably arising from the different nature of the star and gas motions. Usually, 
in a bar-like potential we have several orbit families. For the gas,  being 
collisional, only non-intersecting orbits can exist. In the case of the stars, 
several orbit families can co-exist resulting in a velocity field which has 
isovelocity contours that could be different from those of the gas, and larger 
velocity dispersion (Vauterin \& Dejonghe 1997).
 
\newpage

\begin{figure}
\plotone{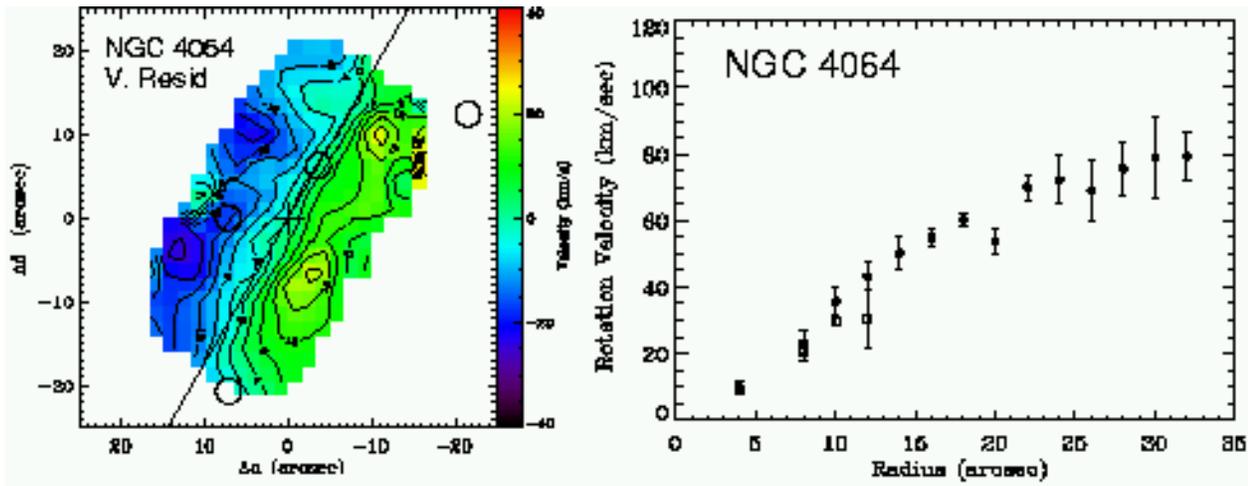}
\caption{NGC 4064 residual stellar velocity map ({\em Left}), and stellar and
ionized gas rotation curve ({\em Right}). Stellar rotation velocities are
represented by solid circles, and H$\beta$ rotation velocities by diamonds.}
\label{n4064rotcurv}

\end{figure}

We now turn to the determination of the stellar rotation curves. Fitting a pure 
circular tilted ring model to the stellar velocity field with 
 fixed P.A = 150$\deg$ and $i=$ 70$\deg$ (Koopmann \etal 2001), 
we found a stellar systemic velocity is 929 $\kms$ $\pm$ 3 $\kms$. The maximum 
rotation velocity over the denspack array is of 80 $\pm$ 7 $\kms$. The residual 
map shows a clear gradient along the minor axis, with a maximum velocity of 
about 20 $\kms$ (Fig $\ref{n4064rotcurv}$).

\begin{figure}

\begin{center}
\epsscale{0.8}
\plotone{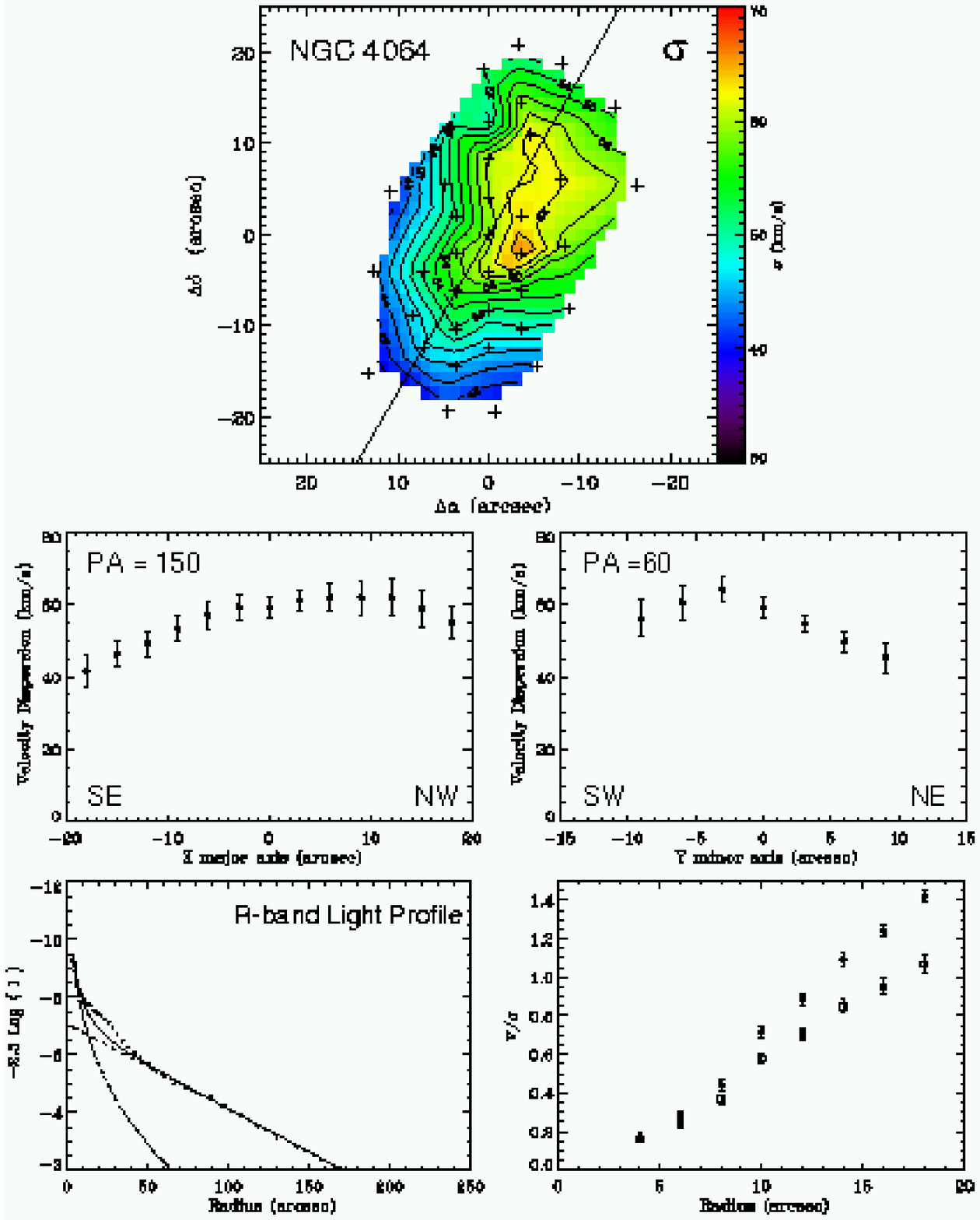}
\end{center}
\caption{NGC 4064 stellar velocity dispersion and $V/\sigma$. {\em Top:}
Stellar velocity dispersion map. {\em Middle Left:} Stellar velocity dispersion
along the optical major axis (P.A = 150$\deg$). {\em Middle Right:} Stellar
velocity dispersion along the optical minor axis (P.A = 60$\deg$).
{\em Bottom Left:} R-band light profile, solid circles represent observed 
light profile, dotted line represents a $R^{1/4}$ component,
dash line represents a exponential disk component, and solid line the 
total light profile. {\em Bottom Right:} $V/\sigma$ as function of the radius. 
Solid circles represent the NW side of the galaxy, and the open circles
represent SE side.}
\label{n4064vdisp}
\end{figure}

The velocity dispersion field has a maximum of $\sim$ 64 $\kms$ with a peak 
slightly
offset from the center. The NW side of the galaxy shows a velocity dispersion 
higher
than the SE side (Fig. $\ref{n4064vdisp}$). Comparing the velocity dispersion
profile with the stellar rotation curve, we find that the galaxy becomes 
supported
by rotation for R $>$ 17" (1.3 kpc) where the $V/\sigma \geq$ 1
(Fig $\ref{n4064vdisp}$) which roughly matches the radius at which the disk begins to dominate the light profile, as expected.

\subsubsubsection{Ionized gas kinematics}

The gas (H$\beta$ and [O III]$\lambda$ 5007 emission), exhibits a kinematical
behavior similar to that of the stars (Fig. $\ref{n4064stellarkin}$). The gas 
emission is confined to the inner 10 arcsec and forms a bar-like structure, with
star formation regions (H$\beta$ emission) along the axis of the bar.
The isovelocity contours do not appear to be aligned with the bar-like structure as clearly as the stellar velocity field or the CO velocity field is.
As we previously did for the stellar velocity field, we determine ionized gas rotation curves by fitting a pure circular, tilted ring model to the H$\beta$ and [O III] velocity field. We found gas systemic velocities of 943 $\pm$ 2 $\kms$, and 947 $\pm$ 2 $\kms$ respectively, which are identical within the errors, but higher than the stellar systemic velocity of 929 $\pm$ 3 $\kms$. 
The ionized gas rotation curves is similar to that of the stars with a maximum velocity of 30 $\pm$ 9 $\kms$ at 12".

\begin{figure}
\epsscale{1.0}
\plotone{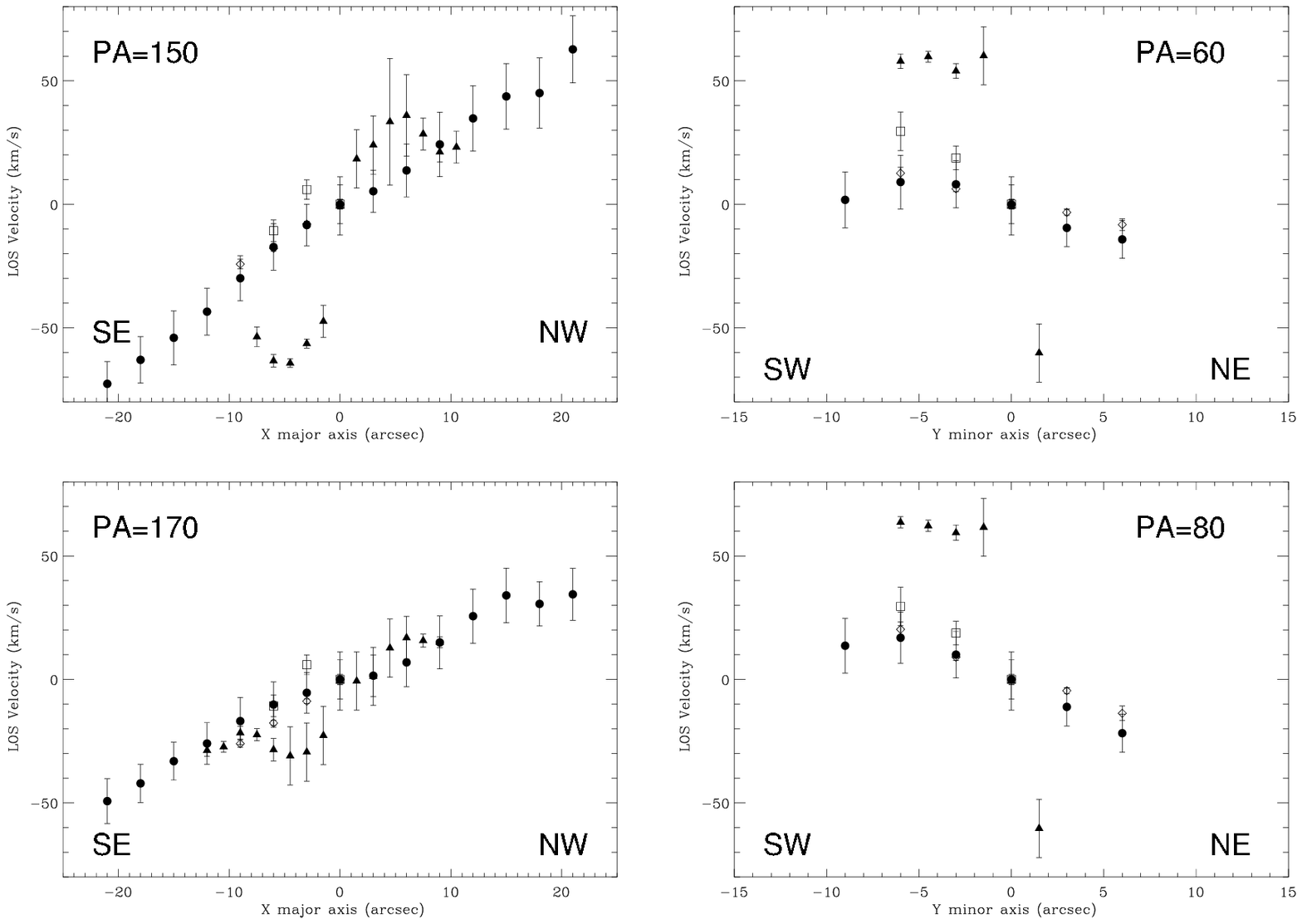}

\caption{Line-of-sight velocities of NGC 4064, along the major and minor axis of 
the outer galaxy (P.A=150$\deg$ and 60$\deg$), and along the major and minor 
axis
 of the bar (P.A=170$\deg$ and 80$\deg$). Stellar velocities are represented by 
solid circle, H$\beta$ velocities by diamonds, [O III] velocities by squares, 
and 
CO velocities by solid triangles.}
\label{n4064cut}
\end{figure}

Along the bar both stellar and ionized gas kinematics are consistent with circular motions as indicated by Fig. $\ref{n4064cut}$. Across the bar stellar and ionized gas velocities exhibit a gradient of about 20 $\kms$, indicating the 
presence of radial streaming motions. A velocity gradient 3 times larger is found for the CO velocities (Fig. $\ref{n4064cut}$) 

\subsubsection{NGC 4424}

\subsubsubsection{Stellar kinematics}

\begin{figure}[ht]
\epsscale{1.0}
\plotone{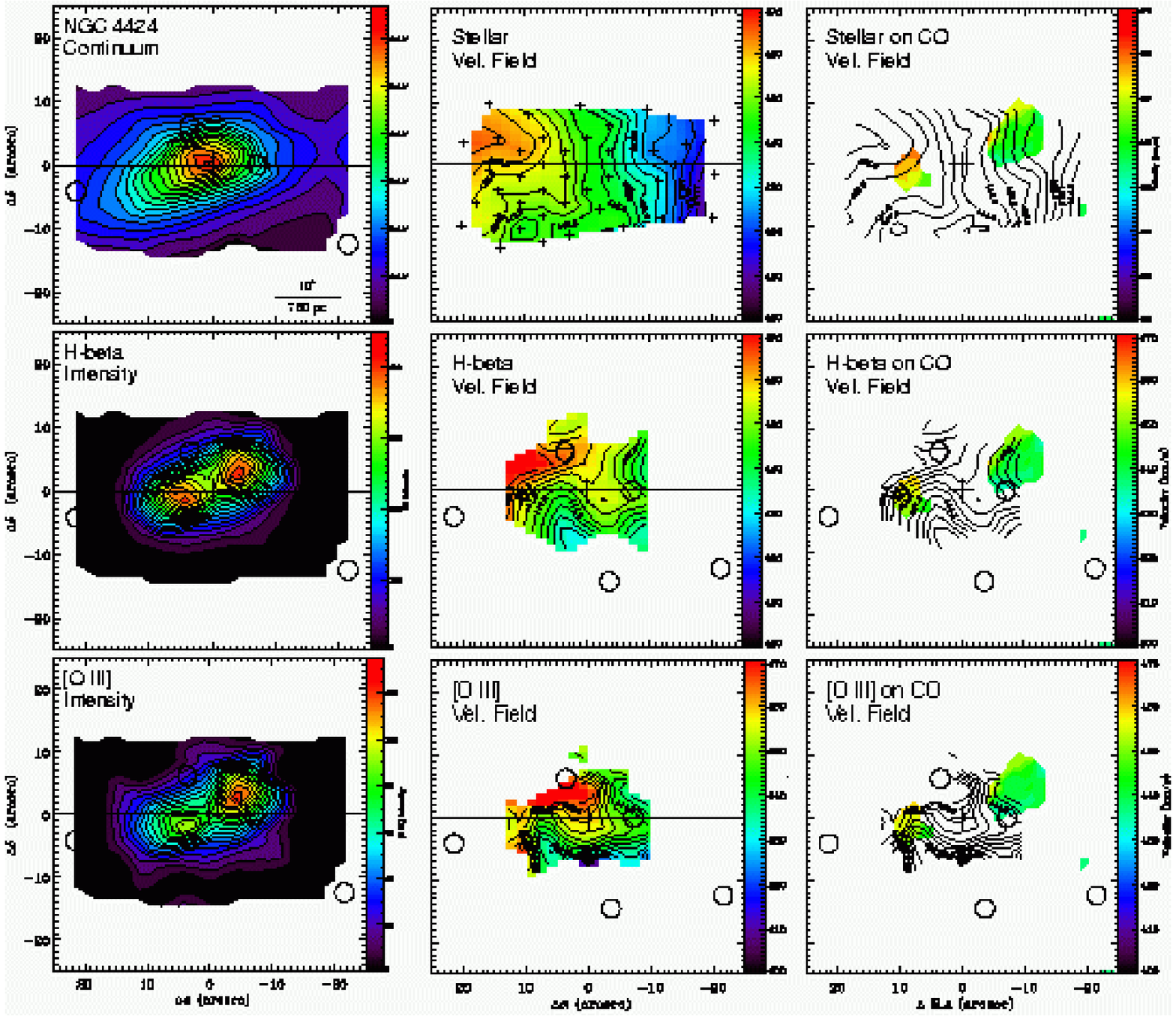}
\caption{DensePak stellar kinematics of NGC 4424. {\em Top Left:} Continuum map. 
{\em Top Middle:} LOS stellar velocity field. {\em Top Right:} LOS stellar 
velocity field (contour lines) over CO velocity field (color). {\em Middle 
Left:} H$\beta$ gas distribution. {\em Middle Middle:} H$\beta$ LOS velocity 
field. {\em Middle Right:} H$\beta$ velocity field (contour lines) over CO 
velocity field (color). {\em Bottom Left:} [O III]$\lambda$5007 gas 
distribution. {\em Bottom Middle:} [O III]$\lambda$5007 velocity field. {\em 
Bottom Right:} [O III]$\lambda$5007 velocity field (contour lines) over CO 
velocity field (color).}

\label{n4424stellarkin}
\end{figure}

The NGC 4424 stellar and ionized gas velocity fields are shown in Fig. $\ref{n4424stellarkin}$. The observed stellar velocity field looks smooth and consistent with circular motions, but with some
distortion in the isovelocity contours probably due to low signal-to-noise ratio.

\begin{figure}
\plotone{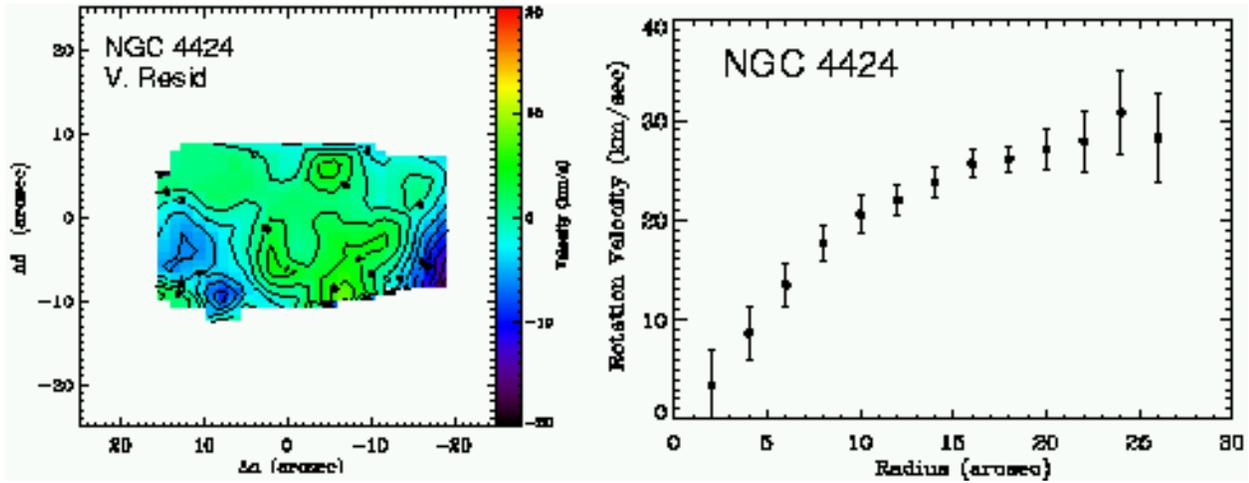}
\caption{NGC 4424 residual stellar velocity map ({\em Left}), and stellar and
ionized gas rotation curve ({\em Right}). Stellar rotation velocity is
represented by solid circles.}
\label{n4424rotcurv}

\end{figure}

As we did for NGC 4064, we fitted a pure circular tilted ring model to the
stellar velocity field with fixed P.A = 90$\deg$ and inclination
$i$ = 60$\deg$ in order to derive the rotation curve. We obtained a smooth
curve with a systemic velocity of 442 $\pm$ 4 $\kms$.
The rotation velocity reaches a maximum of 31 $\pm$ 4 $\kms$ at the end of
the array (Fig. $\ref{n4424rotcurv}$), a very low value for the observed
luminosity but consistent with Rubin et al. (1999).
The residual map between the pure
rotating model and the data reveals small residuals of 4 $\kms$
(Fig. $\ref{n4424rotcurv}$). 
 
\begin{figure}

\newpage 

\begin{center}
\epsscale{0.8}
\plotone{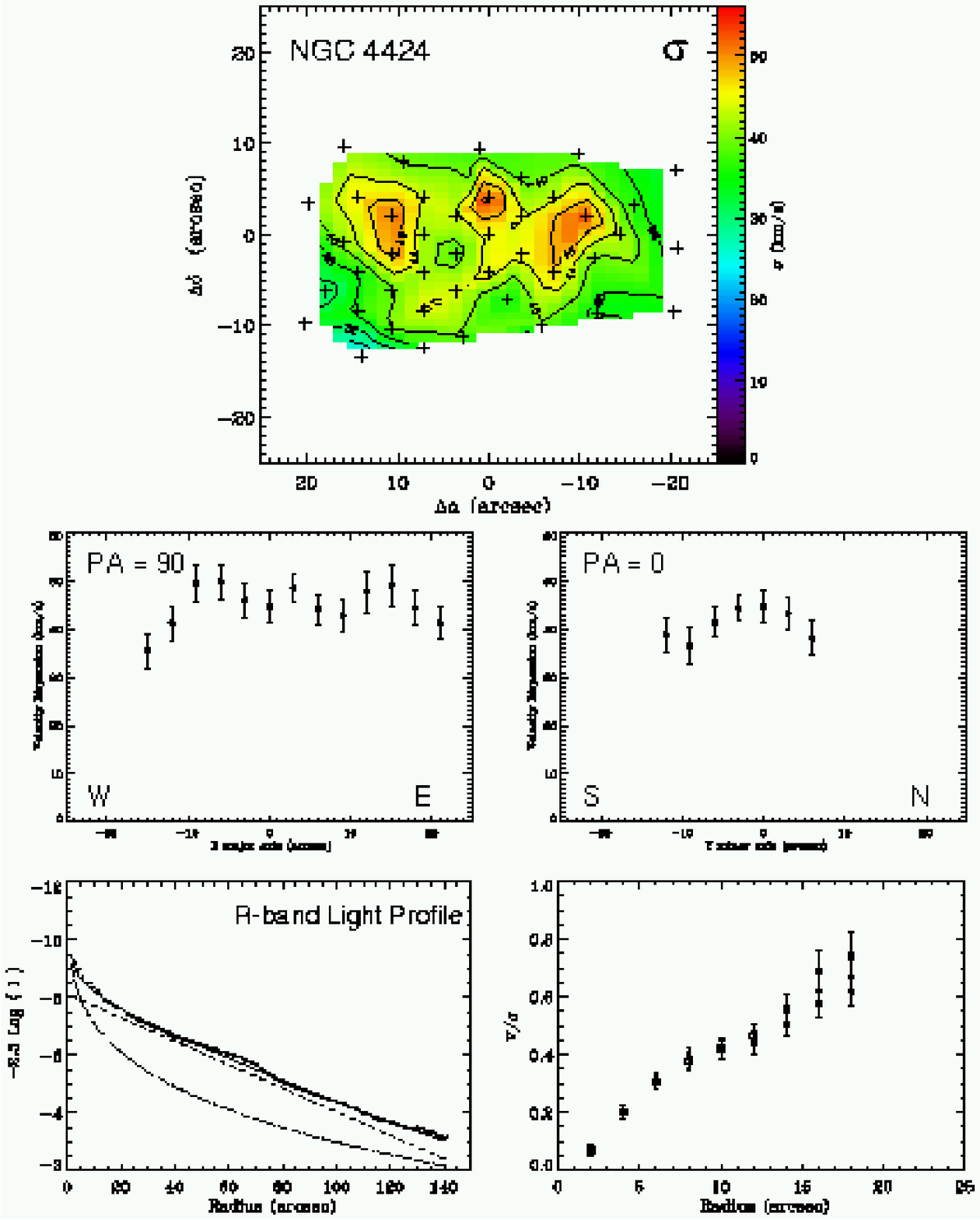}
\end{center}
\caption{NGC 4424 stellar velocity dispersion and $V/\sigma$. {\em Top:}
Stellar velocity dispersion map. {\em Middle Left:} Stellar velocity dispersion
along the optical major axis (P.A = 150$\deg$). {\em Middle Right:} Stellar
velocity dispersion along the optical minor axis (P.A = 60$\deg$).
{\em Bottom Left:} R-band light profile, solid circles represent observed
light profile, dotted line represents a $R^{1/4}$ component,
dash line represents a exponential disk component, and solid line the
total light profile. {\em Bottom Right:} $V/\sigma$ as function of the radius. 
Solid circles represent the NW side of the galaxy, and the open circles
represent SE side. Notice that the galaxy is supported by random motions well inside the disk.}
\label{n4424vdisp}

\end{figure}

The stellar velocity dispersion map shows a central peak of 55 $\pm$ 4 $\kms$ slightly offset by 4" (300 pc) from the center of 
the galaxy (Fig. $\ref{n4424vdisp}$). Secondary  peaks are symmetrically located about 10" (800 pc)from the nucleus. 
Comparing the velocity dispersion profile with the stellar rotation curve, we conclude that the galaxy is dominated, as far as we have measured it, by random motions with a 
$V/\sigma$ value of 0.7  at 18" (Fig. $\ref{n4424vdisp}$), well inside the region of disk dominancy. This is surprising and unlike the normal behavior we found in NGC4064, and can be attributed to disk heating resulting from a merger.

\subsubsubsection{Ionized gas kinematics}

Examination of Fig. $\ref{n4424stellarkin}$ reveals that H$\beta$ and
O[III] emission are confined to two star forming regions roughly 
equidistant from the center ($\sim$ 10'', 780 pc), corresponding 
to the two H$\alpha$ complexes described in \S 4.1.2.

\begin{figure}
\plotone{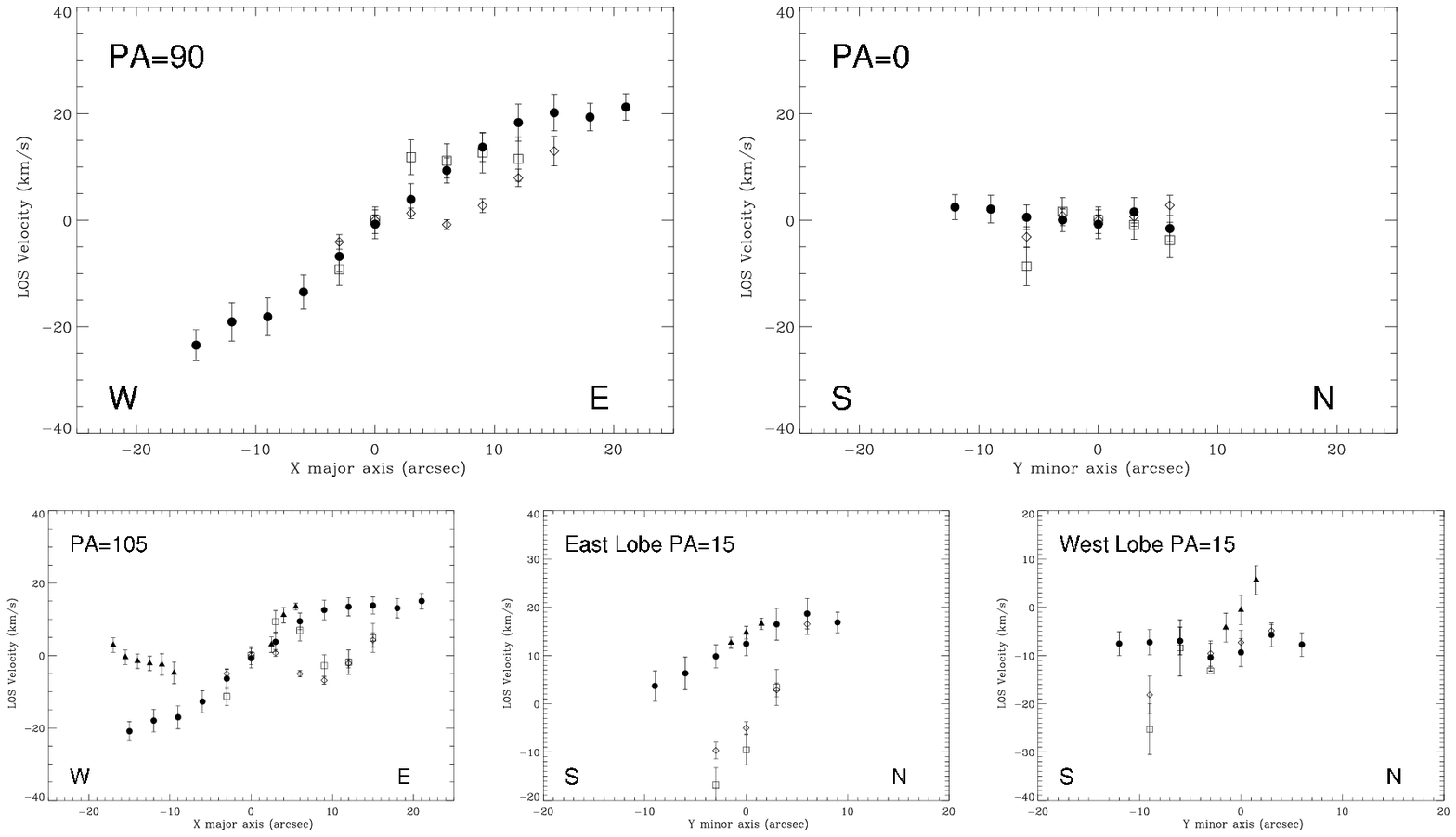}
\caption{Line-of-sight velocities of NGC 4424, along the major and minor axis of
the outer galaxy (P.A=90$\deg$ and 0$\deg$), and along the major and minor axis
 of the lobes (P.A=105$\deg$ and 15$\deg$). Stellar velocities are represented 
by
solid circle, H$\beta$ velocities by diamonds, [O III] velocities by squares, 
and
CO velocities by solid triangles.}
\label{n4424cut}
\end{figure}

The ionized gas velocity field is very different from the stellar 
velocity field. Outside the central 5", 
the gas isovelocity contours tend to be perpendicular to the stellar
isoveocity contours in the outer parts,
indicating that the gas has strong non-circular streaming motions. 
Within the central 15", the H$\beta$ and O[III] velocity fields show a
twisting in the isovelocity contours. Slicing the velocity field along
the major axis (Fig. $\ref{n4424cut}$), and referring to the H$\beta$ velocity field (central panel of Fig. $\ref{n4424stellarkin}$) suggests (within the limits of the Densepak array resolution of 3.5") that inside the Eastern 15" region
the stars and ionized gas have opposite velocity gradients, which might be consistent with gas counterrotation. This is far from conclusive however as the velocity amplitudes are small. The blue-shifted velocity features exhibited by the Southern ionized gas (bottom central panel of Fig. $\ref{n4424stellarkin}$) could be related to the arc-like feature present at the bottom of the H$\alpha$ image in Fig. $\ref{n4424images}$. This might suggest an outflow emerging from the circumnuclear star forming region.

Fig. $\ref{n4424stellarkin}$ shows that the CO behavior is similar to
that of the ionized gas in that both exhibit a velocity gradient along 
the minor axis of the lobes.

\section{Discussion}
\subsection{Origin of the molecular gas bilobal morphology}

NGC 4064 and NGC 4424 each exhibit two distinct concentrations of molecular gas,
or CO lobes. Twin gas concentrations, located symmetrically about the
nucleus are common in the central kiloparsec of galaxies
(e.g. Kenney \etal 1992; Jogee \etal 1999). This is due to the existence of
$m =$ 2 modes, associated with a bar or spiral arms, in galaxy centers. These
$m =$ 2 modes are common especially in barred and interacting systems, and
these concentrate gas into features with $m =$ 2 symmetry. The gas
concentrations are not uniform in radius along these azimuthal features,
because the radial gas flow rate and/or gas consumption rate due to star
formation
is not uniform, leading to gas concentrations at certain radii.

Simulations of gas flow in barred systems (e.g. Athanassoula 1992), show that gas
is driven radially inwards by a combination of gas dissipation and gravitational
torques exerted by the stellar bar.
In simulations without star formation and without a bulge, the
maximum gas concentrations are along the bar and toward the inner parts of the bar.
If there is a bulge, the
maximum gas concentrations are typically deep within the bar, in spiral arm or
ring features, located near kinematically distinct central regions
where $x_{2}$ orbits
(Contopoulos \& Papayannopoulos 1980) begin to occur, and where the potential is
less bar-like and more axisymmetric. Similar behavior may occur in tidally
interacting and merging galaxies, because in tidal interactions there is a
strong  $m =$ 2 perturbation to the gravitational potential, which may cause the
formation of a bar (Barnes \& Hernquist 1996).

The CO distributions in many barred galaxies are more concentrated than the gas
distributions seen in simulations, probably because star formation occurs
preferentially at certain locations, consuming the gas more at some radii than
others. In many
galaxies the star formation occurs preferentially at smaller radii, and the
main gas concentration is at a slightly larger radius
(e.g. Kenney \etal 1992; Kohno \etal 1999)
This suggests either
a time sequence, with radially inflowing gas first concentrating and then
young stars forming, and emerging downstream, and/or a dynamical difference
in the gas at smaller radius making it more susceptible to star formation.
In many galaxies, this combination of effects leads to ``twin peaks" in the
molecular gas distribution.

Are NGC 4064 and NGC 4424 consistent with that picture?
Both galaxies show non-circular motions within the CO lobes,
consistent with bar-like streaming motions, and possibly shocks.
In NGC 4064, the stellar morphology and kinematics both clearly show
the presence of a stellar bar, extending out to a radius of $\sim$30''.
Both galaxies have very small bulges, so
in neither galaxy is there a kinematically distinct central
region observed, as there is in galaxies with larger bulges.
The CO peaks at $r\sim$10'' are well inside the bar,
and are nearly aligned with it.
In NGC 4424, there is no clear evidence of a stellar bar.
The gas bar in NGC~4424 may be a dynamically decoupled gas bar,
which can form in simulations of both mergers (e.g., Barnes \& Hernquist 1996).
and the nuclear regions of barred galaxies
(Englmaier \& Shlosman 2004).
The gas does not presently have enough mass to be self-gravitating
since it comprises only 15$\%$ of the dynamical mass within the central 1 kpc,
but maybe it used to.

Both galaxies have
H$\alpha$ emission located inside the CO lobes, which is consistent with
the picture of a time sequence in the star formation process.
In NGC~4064, the inflow velocity (corrected by inclination) of $\sim$70 $\kms$ and the separation
between the star-forming complexes and CO lobes of about 3" (230 pc)
corresponds to a time delay of $\sim$4 Myrs, which is about the
typical time expected for the onset of star formation.
In NGC~4424, both the inflow velocity\footnotemark[1]\footnotetext[1]{Take it as the inclination corrected velocity along the minor axis of the CO lobes} of 23 $\kms$ and the CO-H$\alpha$
separation of 1"-2" are smaller than in NGC~4064,
but the corresponding time delay is about the same, $\sim$4-8 Myrs.
The consumption of gas in the center by star formation likely explains
why the CO has a bi-lobal rather than bar-like distribution.

\subsection{Star formation history}

NGC 4064 \& NGC 4424 have central star formation rates which
are among the highest in the Virgo Cluster (Koopmann \& Kenney 2004),
and enhanced with respect to typical spiral galaxies, but lower than
extreme starbursts.
Using the H$\alpha$ parameters derived from H$\alpha$ surface photometry
(Koopmann \etal 2001), we find that NGC 4064 has an H$\alpha$ luminosity
of 1.5$\times$10$^{40}$ erg s$^{-1}$, which corresponds to a global star formation rate of 0.12 M$_{\sun}$
yr$^{-1}$ using the prescription given by Kennicutt (1998).
NGC 4424 has a similar H$\alpha$ luminosity
of 1.8$\times$10$^{40}$ erg s$^{-1}$, and a global star formation rate of
0.15 M$_{\sun}$ yr$^{-1}$. These global star formation rates are low, but the
star formation densities in the nucleus are high with a value of
$\sim$ 0.1 M$_{\sun}$ yr$^{-1}$ kpc$^{-2}$.
Both galaxies have total gas masses of 4$\times$10$^{8}$ M$_{\sun}$,
and similar gas consumption timescales of $\sim$3 Gyr.

Optical images show that star formation activity in both galaxies
was more spatially extended in the recent past.
B-R maps (Fig. $\ref{picstarco}$ and Fig. $\ref{n4424picstarco}$)
show blue stellar complexes along the bar in NGC 4064, and along the major axis
for NGC 4424, out to $\sim$30" (2.3 kpc), much further than the present
H$\alpha$ emission.
The blue stellar complexes are within the region
still containing HI (Chung \etal, in prep), suggesting that star formation
has ceased at radii of 10-30$''$ because gravitational torques drove the
gas inwards, rather than the gas being stripped.

\newpage

\subsection{Nature of the NGC 4424 and NGC 4064 galaxies}

NGC 4064 and NGC 4424 show several interesting similarities:
(a) CO and H$\alpha$ gas distributions which are
very truncated and confined to the central kpc.
(b) Bar-like CO and H$\alpha$ morphologies, with two CO peaks located
nearly symmetrically about the
kinematical center, and with star formation regions located inside the CO peaks.
(c) CO distributions which are asymmetric about both the major and minor axes,
suggesting a non-equilibrium gas distribution.
(d) disturbed dust distributions.
Were these galaxies affected by the same
environmental process? Despite these similarities, there are some important
differences between these two objects: First, NGC 4064 shows smooth and
unperturbed outer isophotes (Fig. $\ref{n4064images}$),  consistent
with an unperturbed stellar disk. NGC 4424, on the other hand, shows shell-like
features, and banana-shapes isophotes, indicating a very perturbed stellar disk. We
turn now to the different scenarios which could explain the observed
similarities and differences between the galaxies.

\subsubsection{ ISM-ICM stripping}

NGC 4064 and NGC 4424 are very HI-deficient and have extremely truncated gas disks.
Could gas depletion and truncation be due to ram pressure stripping?
This mechanism is capable of removing the outer gas of cluster spirals, and evidence
shows that it affects mosts spirals in the Virgo cluster
(Giovanelli \& Haynes 1983; Warmels 1988; Cayatte \etal 1990; Solanes \etal 2001).
NGC 4522 is an especially clear example of
HI gas being stripped from the normal stellar disk of a Virgo spiral
(Kenney \etal 2004), and
most Virgo spirals have H$\alpha$ disks which appear spatially truncated
but otherwise normal in their central regions (H$\alpha$/normal),
consistent with ICM-ISM stripping (Koopmann \& Kenney 2004).

%%%%%%%%%%%%%%%%%% paragraph revised by Jeff Kenney %%%%%%%%%%%%%%%%%%%%%%%%
New VLA HI observations (Chung, Van Gorkom, Kenney, \& Vollmer 2005)
show that both galaxies exhibit HI disks truncated at about 5 kpc.
An HI tail with no stellar counterpart
in NGC 4424 suggests that ISM-ICM stripping has affected this galaxy.
NGC 4064 is very far out in the cluster ($d_{proj} \sim $ 2 Mpc),
so it is unlikely that ram pressure stripping is acting now,
unless the ram pressure is much larger than expected from
a static and smooth ICM.
Although it is outside the virial radius of the cluster,
it is probably just inside the maximum rebound radius of the cluster
(Cort\'es \etal in prep.),
making it plausible that it was stripped during a core passage
$\sim$2 Gyr ago.
%%%%%%%%%%%%%%%%%%%%%%%%%%%%%%%%%%%%%%%%%%%%%

Judging from the VLA HI maps, ram
pressure stripping has played a role in the depletion of the gas content in
these galaxies (Chung, Van Gorkom, Kenney, \& Vollmer 2005), but both galaxies
present other features that are not consistent with simple ram pressure
stripping: (1) H$\alpha$ morphologies which are ``truncated/compact'' and
not ``truncated/normal'', as should be the
case in simple ISM-ICM stripping (Koopmann \& Kenney 2004). (2) The central
starburst
suggests that the activity was triggered in the last
few hundred Myr, more recently than a core passage time in the case of NGC 4064.
(3) The disturbed dust distribution is unlikely to be caused by ISM-ICM
stripping, being rather the signature of a recent gravitational interaction
($t \sim $ few orbital times). 4) The disturbed stellar disk in
NGC 4424 cannot be explained by simple ISM-ICM stripping, a mechanism unable to
alter the structure of  stellar disks.
5.) Asymmetric CO distributions in the central kpc suggest a gravitational
disturbance.

All these facts suggest that gravitational
processes have recently acted on both galaxies, perhaps in addition
to ram pressure stripping.
Gravitational interactions may have
played a role in making some gas less bound to the stellar disks, thus
making ram pressure stripping more efficient.

\subsubsection{NGC 4424, a case of minor/intermediate mass merger with ISM-ICM 
stripping}

NGC 4424 has several features that can be associated with a merger event,
including a disturbed stellar distribution with shell-like features and
strong deviations from ellipticity, disturbed dust lanes,
non-circular gas streaming motions, an enhanced central star formation rate,
and ionized gas kinematics in the core which are disturbed and possibly counterrotating.

Simulations show that mergers are very effective in transforming the
morphology of the progenitor galaxies, forming a variety
of objects depending of the mass ratio between the progenitors.
Elliptical galaxies with $r^{1/4}$ radial profiles are the result of
collisions with mass ratio of 1:1 to 3:1 (e.g, Barnes 1992,
Bendo \& Barnes (2000), Cretton \etal 2001, and Naab \& Burkert
2003).  Minor mergers with mass ratios less than 10:1 lead to thickened
spiral disks supported by rotation, and intermediate mass mergers
(1:4 - 1:10 mass ratios) generate  objects with mixed properties such as
disk-like structures with an exponential radial light profile supported by
velocity dispersion rather than rotation (Bournaud \etal 2004).

How does NGC 4424 fit in this picture?
NGC 4424 shows a large fraction of light in non-axisymmetric features which
are short lived, surviving a few rotation periods. These features could be
explained by a recent merger event with a mass ratio $\sim$ 1:10, or an older
remnant with a larger mass ratio. In major mergers, classical tidal tails, which
are not seen in this case, are
formed with the first passage of the companion. This suggests the
intermediate-minor merger nature of this galaxy.
The small bulge or compact stellar component, exponential light profile,
and remarkably low $V/\sigma$ ratio (Fig. $\ref{n4424vdisp}$),
all indicate that this galaxy shares a disk-like morphology with an
elliptical-like kinematics similar to the remnants
produced by Bournaud \etal (2004). This suggest a mass ratio of 1:10-1:4
for the progenitors.

We have also considered whether the properties of NGC 4424 might be
explained by a high--velocity galaxy--galaxy collision, since such
collisions can produce both a gravitational disturbance and ISM--ISM
gas stripping. However, we think that the extent of the disturbances,
especially the banana-shaped stellar distribution, the shell-like
stellar features, and the large velocity dispersion are difficult
to explain through a high-velocity collision, and are more naturally
explained by a merger.

\subsubsection{NGC 4064, case for some type of gravitational interaction plus
gas stripping}

In the case of NGC 4064,
there is evidence for some type of gravitational interaction plus
gas stripping, although the detailed scenario is not as clear as in NGC 4424.
We consider 4 different scenarios.
1. ICM-ISM stripping of a barred spiral galaxy.
2. First ICM-ISM stripping, then a minor merger.
3. First ICM-ISM stripping, then a tidal interaction.
4. A close, high-velocity galaxy-galaxy collision with ISM-ISM stripping.

1.
The simplest scenario is that of a barred spiral galaxy whose outer disk gas
was stripped by ICM-ISM stripping by a passage through the cluster core.
The gravitational torques exerted by the stellar bar
would make some of the remaining gas infall to the center.
If the galaxy has a typical orbital
velocity of 1000 $\kms$, passage through the central 2$\deg$ core,
where most stripping probably occurs, would have occurred $\sim$2 Gyr ago.
This long timescale is difficult to reconcile with
the present central star-forming activity, the disturbed dust
distribution, and the asymmetric CO distribution.
Moreover, the disturbed dust and CO distributions, and the stellar bar itself,
are more naturally explained by a gravitational interaction.

We think the observed properties of NGC~4064 indicate
a combination of gas stripping, either by an ICM-ISM or ISM-ISM interaction,
and a gravitational encounter, either a minor merger or tidal interaction or
direct collision.

2.
Mergers can drive large amounts of gas to the central
kiloparsec ($\sim $ 60 \%, Barnes \& Hernquist, 1996), trigger central
starburts, and produce disturbed
dust distributions. During the first passage, a bar instability develops
due to tidal forces between the progenitors. If the galaxy has a small
bulge or a small dark matter halo which cannot maintain the axial symmetry,
they are more likely to develop a bar instability.
Simulations of collisions of bulgeless galaxies by Iono, Yun, \& Mihos (2004)
show features in the gas morphology which resemble NGC 4064, including
a bar in the center which turns into open spiral arms further out,
and central gas distributions which are asymmetric about the major axis at some
phases.

The lack of H$\alpha$ emission in the outer parts suggest that the galaxy
was depleted of its outer disk gas by ram pressure stripping before the
collision.
It could have looked like NGC 4580, which is also in the outskirts
of the cluster, or like one of the other ``truncated/normal'' galaxies
(Koopmann \& Kenney 2004), before the collision.
A pre-collision gas truncation radius of 0.3 - 0.6 $R_{25}$ seems likely.

A possible difficulty with a merger scenario is that NGC 4064 has an
undisturbed outer stellar disk, without any sign of shells or tails.
This points to a recent minor merger with mass ratio less than 1:10.
The stellar $V/\sigma$ ratio (Fig. $\ref{n4064vdisp}$) reveals an object
largely dominated by rotation, implying that any merger had a small mass ratio.
Even minor mergers can produce
tails and disturbances in the outer disk (e.g, Laine \& Heller, 1999), unless
the collision occurs with a large inclination angle.
Such high inclination encounters are less likely to
disturb the outer disk, and also increase the resettling time of gas onto
the stellar disk, which could explain the persistence of disturbed dust.

3.
An alternate possibility to be considered is that of a non-merging
tidal interaction. Tidal interactions can also
trigger bar instabilities in bulgeless disks, cause central starbursts,
and disturb dust lanes.
The closest candidate for such an interaction is NGC 4049, a small
irregular galaxy, located at a distance of
30' ($\sim$ 130 kpc) from NGC 4064. Its velocity is 829 $\kms$, and its
magnitude is $B=14.25$ which implies a mass of ${\sim}$ $1/6$ the mass of NGC
4064.
The DSS image shows that NGC 4049 is a disturbed galaxy with a bar-like
appearance. If such an encounter took place, it could have happened about 1 Gyr
ago or less.
A possible problem with this scenario is the lack of tidal tails or outer
disk disturbance.
Tidal tails are dispersed more quickly in high-velocity encounters,
and in clusters (Mihos 2004), although the latter effect may not be
so important in the cluster outskirts.
A second possible problem is the disturbed dust lanes. While tidal
interactions can disturb dust lanes somewhat, it is not clear that
they can produce the extent of disturbance seen in NGC~4064.

Minor mergers and non-merging tidal encounters can produce broadly similar
effects, although a high-inclination minor merger may be more able to
produce the observed central disturbances without strongly disturbing the
outer stellar disk.

4.
A close, high-velocity, non-merging, galaxy-galaxy collision
could both disturb the galaxy through gravitational forces,
and strip gas by ram pressure in the ISM-ISM collision.
Such interactions are relatively rare, but offer the potential advantage
of doing everything with only one interaction (Kenney etal 1995).
In this scenario, disturbed dust lanes may arise from gas resettling
into the galaxy after the collision.
However, it is unclear whether the detailed properties of NGC~4064 are
consistent with this scenario. There are no ring-like
or tidal arm disturbances to the disk.
There is no obvious collision partner.
The nearby galaxy NGC 4049 is disturbed, although has more gas than NGC~4064,
inconsistent with a collision scenario since it is a smaller galaxy than
NGC~4064.

Our best scenario for NGC 4064
is a galaxy with a ``truncated/normal'' gas distribution due to a past
ram pressure stripping event now undergoing a minor merger (1:10 - 1:20)
with a large inclination angle. A tidal interaction or collision
with a companion (such as NGC 4049) cannot however be ruled out.

\subsection{ Summary on Environmental Mechanisms and the Future of these Galaxies}

Our results suggest that ''truncated/compact'' galaxies are the result of the
action of at least two environmental processes, namely ram pressure stripping and
gravitational interactions. Ram pressure is most important in the core of
the cluster where the
ICM density is large enough to be efficient in removing the ISM from the
 galaxies. The perturbed morphologies on the other hand are the result of recent
gravitational
interactions which drive gas to the central kiloparsec thus triggering star
formation. This suggests that the morphological transformation of some galaxies
in clusters could be driven by more than a single process.

%%%%%%%%%%%%%%%% paragraph revised by jeff kenney %%%%%%%%%%%%%%%%%%555
The future of these galaxies seems clear. Without any reservoir of gas
left and assuming the present SFR, the galaxies will consume most their
gas
during the next
3 Gyrs. As the gas continues to infall and undergo star formation,
it will form compact circumnuclear stellar disks
(pseudo-bulges)(Kormendy \& Kennicutt 2004),
or perhaps even bulges. Outer stellar structures will fade
in the next Gyr.
Since both galaxies have exponential outer galaxy light profiles,
they would likely end up as S0 galaxies with small but
enhanced bulges or pseudo-bulges. 
 %%%%%%%%%%%%%%%%%%%%%%%%%%%%%%%%%%%%

\section{Summary}
%%%%%%%%%%% Jeff summary %%%%%%%%%%%%%%%%%%%%%%%%%
We have presented optical imaging, 2-D optical spectroscopy and CO
interferometry of the peculiar Virgo cluster galaxies NGC 4064 and NGC
4424.
Our main conclusions are:

 1. Optical imaging reveals that NGC 4424 has a heavily disturbed
stellar disk, with
banana-shaped isophotes and shells. NGC 4064 has a relatively undisturbed outer
stellar
disk, with a central stellar bar that smoothly connects with open
spiral arms in the outer disk.

%%%%%%%%%%%%%% revised by Jeff Kenney %%%%%%%%%%%%%%%%%%%%%%%%
2. Both galaxies show strong H$\alpha$ emission confined to
the central kiloparsec, and little or no H$\alpha$ emission beyond.
H$\alpha$ emission originates from
bar-like strings of luminous star formation complexes, surrounded by
fainter
filaments.
The H$\alpha$ radial distributions,
which resemble the starlight radial distributions of bulges,
have been described as ``truncated/compact''
(Koopmann \& Kenney 2004), and are
distinct from the``truncated/normal'' H$\alpha$ morphology
associated  with simple ICM-ISM stripping that is
found in many Virgo spirals.

%%%%%%%%%%%%%%%%% Point  added by Jeff %%%%%%%%%%%%%%%%%%%%%%%%%%%%%
3. Complexes of young blue stars without associated star formation
exist beyond the present location of ongoing star formation,
indicating a rapidly shrinking star-forming disks.
%%%%%%%%%%%%%%%%%%%%%%%%%%%%%%%%%%%%%%%%%%%%%%%%%%%%

4. The optical B-R images show disturbed dust distributions present in
both galaxies. Clear dust extinction features extend to radii of 2-3
kpc,
much further than the molecular or ionized gas detected by us,
although similar to the blue stellar complexes.
The inner dust morphology appears to be affected
by the active star formation in the central kiloparsec.

5. CO interferometry reveals in each galaxy
two main molecular gas concentrations, located
symmetrically about the nucleus at radii of 500 pc.
The CO lobes have broad, and in some cases, double-peaked line
profiles,
and large velocity gradients,
suggesting the presence of shocks along a bar. In both galaxies,
CO velocity fields show strong non-circular streaming motions,
suggesting that the gas is infalling. H$\alpha$ emission is
radially offset inside the CO lobes, suggesting a time sequence in
which gas is concentrated in a bar, and closer to the center
(downstream)
the gas becomes dense enough to undergo star formation.
Star formation consumes much of the molecular gas in the center,
leaving a bi-lobal rather than bar-like gas distribution.
The CO distributions are asymmetric about both the
major and minor axes,
similar to the gas distributions in some minor merger simulations.

6. In NGC 4064 the stellar velocity field shows
clear kinematic evidence for a strong stellar bar, with
a clear pattern of rotation plus bar-like streaming motions.
There is a velocity gradient along the minor axis
of 20 $\kms$ over 500 pc in the stars,
which contrasts with a much larger value of
70 $\kms$ in the molecular gas.
The stars in the galaxy
are supported predominantly by random motions in
the inner 15", where $V/\sigma \leq$ 1,
and  supported predominantly by rotation by r=20",
where $V/\sigma$=1.4.

7. In NGC 4424, the stellar velocity field shows a pattern of
largely  circular motions, with low line-of-sight velocities.
There is no clear photometric or kinematic evidence of a stellar bar.
The stars in the galaxy are supported predominantly by random motions,
with
$V/\sigma$=0.7 $\pm$ 0.1 at r=18" (1.4 kpc), which is the outermost
point
we have measured.
The ionized gas in the 5$''$ core has disturbed kinematics, and is
possibly counterrotating.

8. With its heavily disturbed stellar morphology, small $V/\sigma$,
disturbed dust lanes, central gas bar, and odd core gas kinematics,
NGC 4424 is very likely an intermediate-mass merger.
The lack of outer disk gas is likely caused by
ram pressure stripping, which would be particularly
effective on gas driven outwards by the merger,
and which might plausibly occur at NGC 4424's present location in the
cluster,
at 0.5 virial radii.

9. The origins of NGC 4064's disturbances are less clear.
Its disturbed dust distribution, asymmetric CO distribution,
strong central bar with central enhancement in star formation
are all suggestive of a recent gravitational encounter,
but one which leaves the outer stellar disk relatively undisturbed.
A likely scenario for the gravitational
encounter is a recent minor merger with a large inclination angle
or a close tidal interaction with the nearby spiral NGC 4049.
The outer disk gas deficiency is likely due to ram pressure stripping,
although its outer cluster location at 1.5 virial radii
means that it was either stripped about 2 Gyr ago in a core crossing,
or stripped more locally and recently
by ram pressure much stronger than the smooth static ICM case.

10. Our guess about the future of these galaxies
is that without any reservoir of gas left and assuming the present SFR,
the galaxies will consume most their remaining gas during the next 3
Gyrs.
As the gas continues to infall and undergo star formation,
it will build compact circumnuclear disks (pseudo-bulges),
and perhaps ultimately (through secular evolution) bulges.
Outer stellar structures will fade in the next Gyr.
Thus both galaxies will end up as S0 galaxies, with smooth stellar
disks,
little gas and dust in the the inner 2 kpc, and
larger (although still small) bulges or pseudo-bulges.

11. While ram pressure stripping likely plays an important role
in the formation of lenticular galaxies in clusters, our results
support the idea that it is not the only mechanism.
Gravitational encounters together with ram pressure
stripping play a key role in forming the most peculiar objects
such as NGC 4424 and NGC 4064. As demonstrated by NGC 4064, some of
these processes can occur in the outskirts of clusters.

\bigskip
\bigskip
\bigskip

We thank to Abhijit Saha for kindly providing IDL routines
for reducing the optical imaging data, Rebecca A. Koopmann for
providing
some of the optical imaging data on NGC 4424, and
Pieter van Dokkum and an anonymous referee for helpful comments which
improved
the paper.
The funding for this research has been provided
by Fundaci\'on Andes Chile, FONDAP project grant 15010003, Chile,
and NSF grant AST-0071251.

\onecolumn

\newpage

\clearpage

\begin{figure}

\plotone{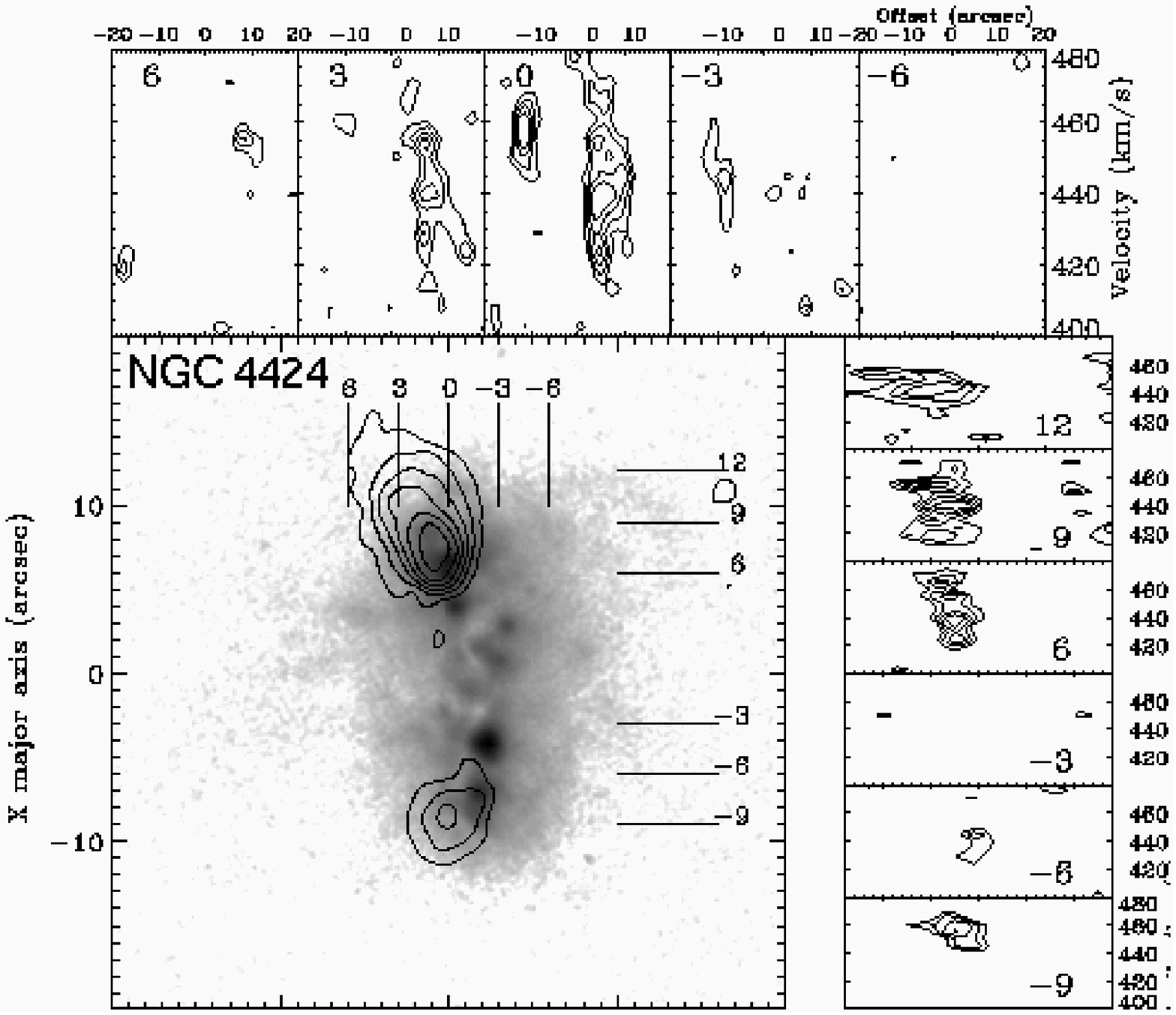}

\bigskip
\bigskip
\caption{
Overlay of CO emission intensity on a gray-scale image of the H$\alpha$ 
 emission of NGC 4424  ({\em left}). CO contours are 1, 2, 3, 4, 5, and 6 
$\times$
 666 mJy beam$^{-1}$ $\kms$. CO position-velocity maps were done with a width of 
1", 
on different
 regions to different
 offset over the major and minor axis of the CO emission (P.A = 105$\deg$). Over 
the top, 
 we have position-velocity diagrams along the major axis of the CO emission, 
showing
 emission spread out over a range of velocities ($\sim$ 40 $\kms$). Over the 
minor axis
 ({\em right}), the velocity gradient is clear, indicating the non-circular 
nature of
 the gas motion.}
\label{n4424pvd}
\end{figure}

\clearpage

\newpage

\clearpage

\clearpage

\newpage

\begin{deluxetable}{cccc}
\tabletypesize{\scriptsize}
\tablewidth{0pc}
\tablecaption{General Properties of NGC 4064 and NGC 4424}
\tablehead{
\colhead{Parameter} & \colhead{NGC 4064} & \colhead{NGC 4424} & \colhead{Note}}
\startdata
Environment & 8.8$\deg$ from M87  & 3$\deg$ from M87  & \\
R.A (J2000) & 12$^{h}$04$^{m}$11.13$^{s}$ & 12$^{h}$27$^{m}$11.6$^{s}$ & \\
Dec (J2000) & 18$\deg$26' 36.84" & 09$\deg$25'14.39" & \\
Morphological type & SBc(s): & Sa pec & 1 \\
    & SB(s)a:pec & SAB0p & 2 \\
Systemic radial velocity (stellar) & 929 $\pm$ 3 $\kms$ & 442 $\pm$ 4 $\kms$ &
\\
Systemic radial velocity (CO) & 929 $\pm$ 5 &  \nodata & \\
Adopted distance & 16 Mpc & 16 Mpc & 3 \\
Linear scale & 77.6 pc arcsec$^{-1}$ & 77.6 pc arcsec$^{-1}$ & \\
B$_{T}^{0}$ & 11.84 & 11.96 & 3 \\
(B-V)$_{T}^{0}$ & 0.68 & 0.62 & 3 \\
M$_{B_{T}^{0}}$ & -19.18 & -19.06 & 3 \\
Optical size & 4.2' $\times$ 1.8' & 3.6' $\times$ 1.9' & 4 \\
R$_{25}$ &138" & 125" & 5 \\
R$_{25}$& 10.7 kpc & 9.7 kpc & 5 \\
P.A (optical isophotes at R$_{25}$) & 150$\deg$ & 90$\deg$ & 5 \\
P.A (CO structures) & 170$\deg$ & 106$\deg$ & \\
Inclination &  70$\deg$ & 61?$\deg$ & 4 \\
HI flux & 1.2 Jy $\kms$ & 2.8 Jy $\kms$ & 6 \\
HI Deficiency & 0.99 & 1.09 & 7 \\
CO flux & 93 $\pm$ 40 Jy $\kms$ & 56 $\pm$ 30 Jy $\kms$ & 8 \\
Maximum stellar rotation velocity & 80 $\kms$ & 31 $\kms$ & 9 \\
FWHM (CO) & 150 $\kms$ & 70 $\kms$ & 8 \\
H$\alpha$ flux & 4.9 $\times$ 10$^{-13}$ ergs cm$^{-2}$ s$^{-1}$ & 6.0 $\times$
10$^{-13}$ ergs cm$^{-2}$ s$^{-1}$ & 5 \\ 
60$\mu$m Flux Density & 3.5 Jy & 3.1 Jy & 10 \\
100$\mu$m Flux Density & 7.1 Jy & 5.8 Jy & 10 \\
T$_{dust}$ & 36 K & 37 K & 10 \\
HI mass & 7.2 $\times$ 10$^{7}$ M$_{\odot}$ & 1.7 $\times$ 10$^{8}$ M$_{\odot}$
& 11 \\
H$_{2}$ mass & 2.6 $\times$ 10$^{8}$ M$_{\odot}$ & 1.6 $\times$ 10$^{8}$
M$_{\odot}$ & 12 \\
M$_{dust}$ & 5.0 $\times$ 10$^{5}$ M$_{\odot}$ & 3.8 $\times$ 10$^{5}$
M$_{\odot}$ & 13 \\
L$_{FIR}$ & 2.2  $\times$ 10$^{9}$ L$_{\odot}$ & 1.8 $\times$ 10$^{9}$
L$_{\odot}$ & 10 \\
L$_{B}$ & 7.3 $\times$ 10$^{9}$ L$_{\odot}$ & 6.2 $\times$ 10$^{9}$ L$_{\odot}$
& 14 \\
\enddata
\tablecomments{(1) Hubble types from Bingelli \etal 1987. (2) Hubble type from RC3.
(3) Jacoby \etal 1992. (4) Optical size at $\mu_{B} \sim $ 25 mag arcsec$^{-2}$.
(5) Radius in units of arc seconds at 25 R mag arcsec$^{-2}$ isophote, Koopmann
\etal 2001. Total H$\alpha$ flux. (6) Helou \etal 1984. (7) Kenney \& Young
1989. The HI
deficiency parameter is defined by Giovanelli \& Haynes 1983. (8) Kenney \&
Young, 1988. (9) this work. Maximum stellar rotation velocity measured
over the Densepak array.
(10) Young \etal 1989. (11) HI mass from $M_{HI}= 2.36 \times 10^{5} D^{2}
S_{HI}$, where $M_{HI}$ is in M$_{\odot}$, $D$ in Mpc, and $S_{HI}$ in Jy $\kms$
(Kenney \& Young 1989). (12) H$_{2}$ mass from $M_{H_{2}}=1.1 \times 10^{4}
D^{2} S_{CO}$, where  $M_{H_{2}}$ is in M$_{\odot}$, $D$ in Mpc, and $S_{CO}$ is
in Jy $\kms$ (Kenney \& Young 1989). (13) Warm dust masses calculated as
described by
Young \etal 1989. (14) Blue luminosity, assuming $M_{B_{\odot}}$ = 5.48, and
corrected for Galactic and internal extinction.}
\end{deluxetable}

\clearpage

\begin{deluxetable}{rccc}
\tabletypesize{\small}
\tablecolumns{4}
\tablewidth{0pc}
\tablecaption{Optical Imaging Observations}
\tablehead{
\colhead{Parameter} & \colhead{NGC 4064} & \multicolumn{2}{c}{NGC 4424}}
\startdata
Date  & 2002 Mar 13 & 2001 Mar 15 & 1997 May 02 \\
Telescope/Chip & WIYN/Mini-mosaic & WIYN/Mini-mosaic & WIYN/S2KB \\
Scale (arcsec pixel$^{-1}$) &  0.141 & 0.141 & 0.20 \\
Filter  & B, V, R, H$\alpha$ (6569/73 {\AA}) &  R, H$\alpha$ (6569/73 {\AA}) &
B, R, V \\
Exposure (sec) & 3 $\times$ 180, 3 $\times$ 180, 3$\times$ 180, 3$\times$ 300 &
3$\times$ 300, 3$\times$ 500 & 3 $\times$ 180, 3 $\times$ 180 \\
FWHM (arcsec) & 0.64, 0.86, 0.54, 0.52 & 0.69, 0.74 & 1.34 \\ 
\enddata
\end{deluxetable}
\clearpage

\begin{deluxetable}{rcc}
\tablecaption{CO Interferometer Observations}
\tablewidth{0pc}
\tablehead{
\colhead{Parameter} & \colhead{NGC 4064} & \colhead{NGC 4424}}
\startdata
Phase center: R.A (J2000) & 12$^{h}$04$^{m}$11.2$^{s}$ &
12$^{h}$27$^{m}$11.5$^{s}$ \\
Dec (J2000) & 18$\deg$ 26'36" & 9$\deg$25 15" \\
Size of synthetized beam & 4".03 $\times$ 3".23 & 4".14 $\times$ 3".32 \\
Linear Size of synthetized beam & 313 $\times$ 251 pc & 321 $\times$ 258 pc \\
Beam Position angle & -71$\deg$ & -81$\deg$ \\
Number of channels & 120 & 120 \\
Velocity resolution (per channel) & 10.4 $\kms$ & 5.2 $\kms$ \\
Velocity coverage of the emission & 843 to 1021 $\kms$ & 424 to 466 $\kms$ \\
RMS noise in channel maps & 0.015 Jy beam$^{-1}$ & 0.020 Jy beam$^{-1}$ \\
Interferometric flux & 61 $\pm$ 13 Jy $\kms$ & 14 $\pm$ 1 Jy $\kms$ \\ 
\enddata
\end{deluxetable}
\clearpage

\begin{deluxetable}{rc}
\tablewidth{0pc}
\tablecaption{Instrumental Setup for DensePak observations}
\tablehead{
\colhead{Parameter} & \colhead{} }
\startdata
Spatial sampling & 4" (310 pc)   \\
Field-of-View & 30"$\times$45" (2.3$\times$3.5 kpc)  \\
Spectral sampling & 0.48 {\AA} per pixel  \\
Spectral resolution & 2.02 {\AA}  \\
Wavelength range & 4500 - 5500 {\AA}  \\
Exposure & 4$\times$1800 sec  \\
%P.A of array & 150$\deg$ & 90$\deg$ \\
Template star & HD 90861 \\ 
\enddata
\end{deluxetable}


\begin{references}
\reference{} Athanassoula, E. 1992, MNRAS, 259, 345
\reference{} Balogh, M. L., Schade, D., Morris, S. L., Yee, H. K. C.,
 Carlberg, R. G., Ellingson, E. 1998, ApJ, 504, 75
\reference{} Barden, S. C., Sawyer, D. G., \& Honeycutt, R. K. 1998, Proc. SPIE, 3355, 892
\reference{} Barnes, J. E. 1992, ApJ, 393, 484
\reference{} Barnes, J. E., \& Hernquist, L., 1996, ApJ, 471, 115
\reference{} Bendo, G. J., \& Barnes, J. E. 2000, MNRAS, 316, 315
\reference{} Begeman, K. G. 1989, A\&A, 223, 47
\reference{} Bingelli, B., Tammann, G. A., \& Sandage, A. 1987, AJ, 94, 241
\reference{} Binney, J., \& Tremaine, S. 1987, Galactic Dynamics (Princeton: 
Princeton
 Univ. Press)
\reference{} Bloemen, J. B. G. M., \etal 1986, A\&A, 154, 25
\reference{} Bournaud, F., Combes, F., \& Jog, C. J. 2004, A\&A 418, L27 
\reference{} Butcher, H., \& Oemler, A. Jr. 1978, ApJ, 219, 18
\reference{} Cappellari M., Copin Y. 2003, MNRAS, 342, 345
\reference{} Cayatte, V., van Gorkom, J. H., Balkowski, C., \& Kotanyi, C. 1990, AJ, 100, 604
\reference{} Chung, A., van Gorkom, J. H., Kenney, J. D. P., \& Vollmer, B. 2005, in ``Extraplanar Gas",
ed R. Braun, ASP Conf Series, Vol. 331, p 275--280
\reference{} Contopoulos, G., \& Papayannopoulos, T. 1980, A\&A, 92, 33
\reference{} Cretton, N., Naab, T., Rix, H.-W., \& Burkert, A. 2001, ApJ, 597, 
893
\reference{} deVaucouleurs, G., deVaucouleurs, A., Corwin, H. G., Buta, R. J., 
Paturel, G., Fouqu\'e, P.
1991, {\em Third Reference Catalog of Bright Galaxies}, (New York: Springer-
Verlag)
\reference{} Dressler, A. 1980, 236, 351
\reference{} Dressler, A., Oemler, A. Jr., Couch, W. J., Smail, I., Ellis, R. S., Barger, A.,
Butcher, H., Poggianti, B. M., \& Sharples, R. M. 1997, ApJ, 490, 577
\reference{} Englmaier, P., \&  Shlosman, I. 2004, ApJ, 617, L115
\reference{} Freedman, W. L., \etal 1994, Nature, 371, 757
\reference{} Ghigna, S., Moore, B., Governato, F., Lake, G., Quinn, T., \& Stadel, J. 1998, MNRAS, 300, 146
\reference{} Giovanelli, R., \& Haynes, M. P. 1983, AJ, 88, 881
\reference{} G\'omez, P. L. \etal, 2003, ApJ, 584, 210 (17 authors) 
\reference{} Gunn, J. E., \& Gott, J. R. 1972, ApJ, 176, 1
\reference{} Helou, G., Hoffman, G. L., \& Salpeter, e. E. 1984, ApJS, 55, 433
\reference{} Hernquist, L. 1992, ApJ, 400, 460
\reference{} Hubble, E., \& Humason, M. L. 1931, ApJ, 74, 43
\reference{} Iono, D., Yun, M. S., \& Mihos j. C. 2004, ApJ, 616, 199
\reference{} Jacoby, G. H., \etal 1992, PASP, 104, 599
\reference{} Jogee, S., Kenney, J. D. P., \& Smith, B. J. 1999, ApJ, 526, 665
\reference{} Jogee, S. \etal 2004, ApJ, 615, 105
\reference{} Kenney, J. D. P., \& Young, J. S. 1988, ApJS, 66, 261
\reference{} Kenney, J. D. P., \& Young, J. S. 1989, ApJ, 344, 171
\reference{} Kenney, J. D. P., Wilson, C., D., Scoville, N.Z., Devereux, N. A., 
\& Young, J. S. 1992, ApJ, 395, 179
\reference{} Kenney, J. D. P., Rubin, V. C., Planesas, P., \& Young, J. S. 1995, ApJ, 438, 135
\reference{} Kenney, J. D. P., \& Koopmann, R. A., Rubin, V. C., \& Young, J. S. 
1996, AJ, 111, 152
\reference{} Kenney, J. D. P., van Gorkom, J. H., \& Vollmer, B. 2004, AJ, 127, 3361
\reference{} Kennicutt, R. C., Jr. 1998, ARA\&A, 36, 189
\reference{} Kohno, K., Kawabe, R., \& Vila-Vilar\'o, B. 1999, ApJ, 511, 157

\reference{} Koopmann, R. A., Kenney, J. D. P., \& Young, J. 2001, ApJS, 135, 
125
\reference{} Koopmann, R. A., \& Kenney, J. D. P. 2004, ApJ, 613, 866
\reference{} Laine, S., Kenney, J. D. P., Yun, M. S., \& Gottesman, S. T. 1999,
ApJ, 511, 709
\reference{} Laine, S., \& Heller, C. H. 1999, MNRAS, 308, 557
\reference{} Lewis, I., etal 2002, MNRAS, 334, 673  (24 authors)
\reference{} Mastropietro, C., Moore, B., Diemand, J., Mayer, L., \& Stadel, J.
2004, in Outskirts of Galaxy Clusters: Intense Life in the Suburbs, Proceedings of IAU Symposium, No. 222. (Cambridge: Cambridge University Press), 519
\reference{} Mihos, J. C. 2004, in Clusters of Galaxies: Porbes of Cosmological Structure and Galaxy Evolution, ed. J. S. Mulchaey, A. Dressler, \& A. Oemler (Cambridge: Cambridge Univ. Press), 278 
\reference{} Moore, B., Katz, N., Lake, G., Dressler, A. \& Oemler, A. 1996, 
Nature, 379, 613
\reference{} Moore, B., Lake, G., Quinn, T. \& Stadel, J. 1999, MNRAS, 304, 465
\reference{} Naab, T., \& Burkert, A. 2003, ApJ, 597, 893
\reference{} Nilson, P. 1973, Uppsala General Catalogue of Galaxies (Uppsala)
\reference{} Nichol, R. C. 2004,  in ``Cluster of Galaxies: Probes of Cosmological
 Structure and Galaxy Evolution'', ed J. S. Mulchaey, A. Dressler, \& A. Oemler
(Cambridge: Cambridge Univ. Press), 24
\reference{} Nulse, P. E. J. 1982, MNRAS, 198, 1007
\reference{} Poggianti, B. M., Smail, I., Dressler, A., Couch, W. J., Barger, A. J., Butcher, H., Ellis, R. S., Oemler, A. Jr. 1999, ApJ, 518, 576
\reference{}  Rubin, V. C., Waterman, A. H., \& Kenney, J. D. P. 1999, AJ, 118, 236
\reference{} Saha, A., Armandroff, T., Sawyer, D. G., \& Corson, C. 2000, Proc. 
SPIE, 4008, 447
\reference{} 
       Scoville, N. Z., Carlstrom, J. E., Chandler, C. J., Phillips, J. A., 
Scott, S. L., Tilanus,
       R. P. J., \& Wang, Z. 1993, PASP, 105, 1482
\reference{} Schulz, S. \& Struck, C. 2001, MNRAS, 328, 185
\reference{} Solanes, J. M., Manrique, A., Garc\'{\i}a-G\'omez, C., Gonz\'alez-Casado, G., Giovanelli, R., \& Haynes, M. P. 2001, ApJ, 548, 97
\reference{} Spitzer, L. Jr., \& Baade, W. 1951, ApJ, 113, 413
\reference{} Statler, T. S. 1995, AJ, 109, 1371
\reference{} Tonry, J., \& Davis, M. 1979, 84, 1511
\reference{} Valdes, F. 1995, {\em Guide to the HYDRA reduction task DOHYDRA} 
\reference{} van der Marel, R. P., \& Franx M. 1993, ApJ, 407, 525
\reference{} van Dokkum, P. G., Franx, M., Fabricant, D., Kelson, D. D., Illingworth, G. D. 1999, ApJ, 520, L95
\reference{} van Gorkom, J. H. 2004, in ``Cluster of Galaxies: Probes of Cosmological
 Structure and Galaxy Evolution'', ed J. S. Mulchaey, A. Dressler, \& A. Oemler 
(Cambridge: Cambridge Univ. Press), 306.
\reference{} Vauterin, P., \& Dejonghe, H. 1997, MNRAS, 286, 812
\reference{} Vollmer, B., Cayatte, V., Balkowski, C., \& Duschl, W., J. 2001, 
ApJ, 561, 708
\reference{} Warmels, R. H. 1988, A\&AS, 72, 19
\reference{} Young, J. S., Xie, S., Kenney, J. D. P., \& Rice, W. L. 1989, ApJS, 
70, 699
\end{references}
\end{document}